\documentclass[ALICE,manyauthors]{cernphprep}

\def\pt {\mbox{$p_{\rm T}$}\xspace}   
 
\def\gevc {\mbox{GeV/$c$}\xspace}

\def\ppb {p--Pb\xspace}
\def\pbp {Pb--p\xspace}

\def\pp {pp\xspace}
\def\jpsi {\mbox{J/$\psi$}\xspace}
\def\psip {\mbox{$\psi$(2S)}\xspace}

\newcommand {\sqrtSnn}   {\ensuremath{\sqrt{s_{\mathrm{NN}}}}}

\def\sigss {\mbox{$\sigma_{\psi({\rm 2S})}$}\xspace}
\def\sigs {\mbox{$\sigma_{{\rm J/}\psi}$}\xspace}

\usepackage[T1]{fontenc}
\usepackage{datetime}
\usepackage{amsmath}
\usepackage{xcolor}
\usepackage[comma,square,numbers,sort&compress]{natbib}
\usepackage{hyperref}
\usepackage{lineno}
\usepackage{multirow}
\usepackage{gensymb}

\begin{document}%

\begin{titlepage}
\PHyear{2020}
\PHnumber{036}       
\PHdate{10 March}  

%

\title{Measurement of nuclear effects on \psip production \\ in \ppb collisions  at  $\sqrt{\textit{s}_{\rm NN}} = 8.16$~TeV}
\ShortTitle{Measurement of nuclear effects on $\psip$ in \ppb collisions at  $\sqrt{\textit{s}_{\rm NN}} = 8.16$~TeV }  

\Collaboration{ALICE Collaboration\thanks{See Appendix~\ref{app:collab} for the list of collaboration members}}
\ShortAuthor{ALICE Collaboration} 


\begin{abstract}
Inclusive \psip production is measured in \ppb collisions at the centre-of-mass energy per nucleon--nucleon pair $\sqrt{s_{_{\rm NN}}}=8.16$ TeV, using the ALICE detector at the CERN LHC.  
The production of \psip is studied at forward ($2.03 < y_{\rm cms} < 3.53$) and backward ($-4.46 < y_{\rm cms} < -2.96$) centre-of-mass rapidity and for transverse momentum \pt $<$ 12 \gevc via the decay to muon pairs. 
In this paper, we report the integrated as well as the $y_{\rm cms}$- and $p_{\rm T}$-differential inclusive production cross sections. 
Nuclear effects on \psip production are studied via the determination of the nuclear modification factor that shows a strong suppression at both forward and backward centre-of-mass rapidities. Comparisons with corresponding results for inclusive \jpsi show a similar suppression for the two states at forward rapidity (p-going direction), but a stronger suppression for \psip at backward rapidity (Pb-going direction). As a function of $p_{\rm T}$, no clear dependence of the nuclear modification factor is found. 
The relative size of nuclear effects on \psip production compared to \jpsi is also  studied via the double ratio of production cross sections $[\sigss/\sigs]_{\rm pPb}/[\sigss/\sigs]_{\rm pp}$ between \ppb and \pp collisions.  
The results are compared with theoretical models that include various effects related to the initial and final state of the collision system and also with previous measurements at $\sqrt{s_{_{\rm NN}}}=5.02$ TeV.

\end{abstract}
\end{titlepage}
\setcounter{page}{2}
%
%

\section{Introduction}

The study of charmonia, bound states of charm (c) and anticharm ($\overline{\rm c}$) quarks, is an important and interesting research domain.
High-energy \pp collisions provide a testground to apply quantum chromodynamics (QCD) theory for understanding the charmonium production mechanism.
The production of heavy-quark pairs, ${\rm c}\overline{\rm c}$ in the present case, is an inherently perturbative process since the momentum transfer is at least as large as the heavy-quark pair mass. On the contrary, the formation of the bound state is achieved on a longer time scale and thus has to be considered as a non-perturbative process. QCD-based approaches such as Non-Relativistic QCD (NRQCD)~\cite{Bodwin:1994jh} give a good description of the main features of quarkonium production cross sections in pp collisions. 
When the production of heavy quarkonium occurs inside a medium, as it happens in case of heavy-ion collisions, it is influenced by the properties of the medium and various effects are present.
They are mainly categorised in two groups, hot matter effects and cold nuclear matter (CNM) effects.
Among the former, those related to the formation of a Quark--Gluon Plasma (QGP),
a high energy-density medium created in ultra-relativistic heavy-ion collisions where quarks and gluons are deconfined,  are currently scrutinised at collider experiments at RHIC (mainly Au--Au)~\cite{Adam:2019rbk}, up to $\sqrt{s_{\rm NN}}=0.2$ TeV and the LHC (mainly Pb--Pb)~\cite{Adam:2016rdg,Acharya:2019lkh,Sirunyan:2017isk,Aaboud:2018quy}, up to $\sqrt{s_{\rm NN}}=5.02$ TeV. 
For the \jpsi (1S state with J$^{\rm PC}=1^{--}$), a reduced production with respect to pp collisions was reported, ascribed to dissociation in the QGP as a result of color Debye screening~\cite{Matsui:1986dk}. However, LHC experiments reported a significantly reduced suppression for \jpsi with respect to RHIC, now commonly ascribed to a recombination mechanism~\cite{BraunMunzinger:2000px,Thews:2000rj} related to the much larger multiplicity of charm quarks observed at the LHC~\cite{Adam:2016ich}.
When considering the weakly bound \psip state, Debye screening should lead to a stronger suppression, which at the same time could be influenced by recombination effects. Results currently available at LHC energies on the relative suppression of \psip and \jpsi~\cite{Khachatryan:2014bva,Sirunyan:2016znt,Adam:2015isa} generally show a stronger effect for the former, except for CMS data on Pb--Pb collisions at $\sqrt{s_{\rm NN}}=2.76$ TeV in the kinematic window $3<p_{\rm T}<30$ GeV/c, $1.6<|y|<2.4$ where the opposite behaviour was found. Attempts to explain these observations were carried out~\cite{Du:2015wha}, and it is generally recognised that further precision measurements are needed and might help reaching a final assessment~\cite{Dainese:2703572}.

In addition to more accurate data, a quantitative understanding of the results requires the evaluation of the size of CNM effects, since those are also present in heavy-ion collisions. Among these effects an important role is played by nuclear shadowing~\cite{Eskola:2016oht}, the modification of the partonic structure functions inside nuclei. It leads to a change in the probability for a 
quark or gluon to carry a fraction $x$ of the nucleon momentum and, as a consequence, it affects the production cross section of the c${\rm\overline c}$ pair. At low $x$, this effect could originate from the formation of a 
Color Glass Condensate (CGC)~\cite{Gelis:2010nm}, which can happen when, at high energy, the density of low-$x$ quarks and gluons becomes very large, leading to saturation effects. 
A further mechanism which can also modify the parton kinematics is coherent energy loss, an effect involving partons in the initial and final state~\cite{Arleo:2012hn}. Finally,  hadronic/nuclear break-up of the final-state ${\rm c}{\overline {\rm c}}$ pair~\cite{McGlinchey:2012bp} can also occur, and leads to suppression effects. The common way to investigate CNM effects  is via proton--nucleus collisions, where hot-matter effects are, in principle, negligible.


Various results on CNM effects on charmonium production are available at LHC energies for \ppb collisions at $\sqrt{s_{\rm NN}}=5.02$ TeV. For  J/$\psi$, extensive studies were performed at forward/backward centre-of-mass rapidity $y_{\rm cms}$ by ALICE ~\cite{Abelev:2013yxa,Adam:2015jsa,Adam:2015iga,Adamova:2017uhu} and  LHCb~\cite{Aaij:2013zxa}, as well as at midrapidity by ALICE~\cite{Adam:2015iga}, ATLAS~\cite{Aad:2015ddl} and CMS~\cite{Sirunyan:2017mzd}. A general feature of the results is the observation of a significant \jpsi suppression at forward $y_{\rm cms}$ (p-going direction), which becomes weaker and finally disappears moving towards backward rapidity (Pb-going direction). Theory models which include shadowing effects based on various parameterizations of the nuclear modifications of parton distribution functions are able to reproduce the results~\cite{Albacete:2013ei,Lansberg:2016deg}. At the same time, also models based on a CGC approach~\cite{Ma:2015sia}, or including coherent energy loss as a main CNM mechanism~\cite{Arleo:2013zua}, are in good agreement with data. Such  an agreement with the models described above also implies that the presence of significant break-up effects of the c${\rm\overline c}$ pair, which are not included in these models, is disfavoured.

For \psip, results at $\sqrt{s_{\rm NN}}=5.02$ TeV~\cite{Abelev:2014zpa,Adam:2016ohd,Aaij:2016eyl,
Sirunyan:2018pse,Aaboud:2017cif} clearly showed a larger suppression with respect to \jpsi, in particular at backward rapidity. The CNM effects mentioned in the previous paragraph in conjunction with \jpsi results are initial-state effects or anyway directly related to the hard production of the heavy-quark pair, and are expected to affect similarly both charmonium final states. The additional suppression exhibited by the \psip was therefore attributed to a break-up of this more loosely bound state via collisions with the dense system of interacting particles produced in \ppb  collision~\cite{Ma:2017rsu,Ferreiro:2014bia,Du:2015wha}. It has to be noted that a similar effect was observed, although with larger uncertainties, by the PHENIX experiment in \mbox{p-Al} and \mbox{p-Au} collisions at $\sqrt{s_{\rm NN}}=0.2$ TeV~\cite{Adare:2016psx}.

More recently, with the start of LHC Run 2, \ppb collisions at $\sqrt{s_{\rm NN}}=8.16$ TeV became available. First results on \jpsi, obtained by ALICE~\cite{Acharya:2018kxc} and LHCb~\cite{Aaij:2017cqq}, were compatible within uncertainties with those obtained at $\sqrt{s_{\rm NN}}=5.02$ TeV. In this paper, we show the first results on inclusive \psip production in \ppb collision at $\sqrt{s_{\rm NN}}=8.16$ TeV. Section 2 provides a short description of the apparatus and event selection criteria, while the data analysis for \psip production is described in Sect.~3. Section 4 contains the results, with model comparisons and discussion, and finally a short summary is given in Sect.~5.

%
%

\section{Experimental apparatus and event selection}

Extensive descriptions of the ALICE apparatus and its performance can be found in Refs.~\cite{Aamodt:2008zz,Abelev:2014ffa}.
The analysis presented in this paper is based on muons detected at forward rapidity with the muon spectrometer~\cite{Aamodt:2011gj}. The spectrometer covers the pseudo-rapidity range $-4<\eta_{\mathrm{lab}}<-2.5$ and includes five tracking stations (Cathode Pad Chambers), the central one embedded inside a dipole magnet with a 3 $\mathrm{T \cdot m}$ field integral. Each tracking station consists of two tracking chambers aimed at measuring muons in the bending (vertical) and non-bending (horizontal) planes. Two trigger stations (Resistive Plate Chambers), positioned downstream of the tracking system, provide a single muon as well as a dimuon trigger, with a programmable muon \pt threshold that was set to 0.5 GeV/$c$ for this data sample. An absorber, made of concrete, carbon and steel (with a thickness of 10 interaction lengths) is positioned in front of the tracking system, to remove hadrons produced at the interaction vertex. Hadrons which escape this front absorber are further filtered out by a second absorber, placed between the tracking and the triggering system, which also removes low-momentum muons originating from pion and kaon decays, thereby reducing the background. The position of the interaction vertex is determined by the two layers of the Silicon Pixel Detector (SPD)~\cite{Aamodt:2010aa}, corresponding to the inner part of the ALICE Inner Tracking System (ITS), which  cover the pseudo-rapidity intervals $|\eta_{\rm lab}|<2$ and $|\eta_{\rm lab}|<1.4$. The V0 detector~\cite{Abbas:2013taa}, composed of scintillators located at both sides of the interaction point, and covering the pseudo-rapidity intervals $-3.7<\eta_{\mathrm{lab}}<-1.7$ and $2.8<\eta_{\mathrm{lab}}<5.1$, provides the minimum bias trigger. In addition, the V0 is used for luminosity determination, which is also independently estimated by means of the two T0 Cherenkov detectors~\cite{Bondila:2005xy}, which cover the pseudo-rapidity intervals $-3.3<\eta_{\mathrm{lab}}<-3.0$ and  $4.6<\eta_{\mathrm{lab}}<4.9$. 

The data samples were collected with two different beam configurations, which correspond to the acceptance regions  $2.03<y_{\mathrm{cms}}<3.53$ and $-4.46<y_{\mathrm{cms}}<-2.96$ for dimuons. These configurations were obtained by reversing the direction of the two beams, and are respectively named \ppb (forward) and \pbp (backward) in the following. Positive rapidities correspond to the situation where the proton beam travels towards the muon spectrometer. The integrated luminosities collected for the two configurations are $\textit{L}_{\mathrm{int}}=8.4\pm0.2\;\mathrm{nb^{-1}}$ for \ppb  and $\textit{L}_{\mathrm{int}}=12.8\pm0.3\;\mathrm{nb^{-1}}$ for \pbp collisions~\cite{ALICE-PUBLIC-2018-002}.

Events selected for this analysis were collected by requiring a coincidence between the minimum bias and the dimuon trigger conditions. In order to reject tracks at the edge of the spectrometer acceptance, the pseudo-rapidity selection  $-4<\eta_{\mu,{\rm lab}}<-2.5$ is performed while, to remove tracks crossing the denser regions of the absorber, their radial transverse position ($R_{\mathrm{abs}}$) at the end of the absorber must be in the range $17.6<R_{\mathrm{abs}}<89.5\;\mathrm{cm}$. Finally, the matching based on a $\chi^2$ minimization algorithm between a track in the tracking chambers and a track reconstructed in the trigger system is required.
 

\section{Data analysis}\label{section:analysis}
The analysis procedure reported here is similar to the one discussed in Refs.~\cite{Abelev:2014zpa,Acharya:2018kxc}.
The cross section for inclusive \psip production times the branching ratio B.R.$_{\psi(2{\rm S})\rightarrow\mu^+\mu^-} = (0.80 \pm 0.06)$\%~\cite{Tanabashi:2018oca} is given by

\begin{equation}
{\rm B.R.}_{\psi(2{\rm S})\rightarrow\mu^+\mu^-}\cdot\frac{{\rm d^{2}}\sigma^{\psi(2{\rm S})}_{\rm{pPb}}}{{\rm d}p_{\rm{T}}{\rm d}y} = \frac{N_{\psi(2{\rm S})}^{\rm corr}(y,\pt)}{L_{\rm int}\cdot\Delta y\Delta \pt}
\end{equation}

where $N_{\psi(2{\rm S})}^{{\rm corr}}(y,\pt)$ is the number of \psip in the corresponding $y$ and $\pt$ interval, corrected by the product of acceptance times reconstruction efficiency $A\cdot\epsilon(y,\pt)$, $L_{\rm int}$ is the integrated luminosity and $\Delta y$, $\Delta p_{\rm T}$ are the width of the rapidity and transverse momentum intervals. The choice of not correcting for the decay branching ratio is due to the non-negligible systematic uncertainty it would introduce ($\sim$8\%~\cite{Tanabashi:2018oca}).

 The number of reconstructed J/$\psi$ and \psip resonances are extracted via fits to the invariant mass spectrum of opposite-sign muon pairs. More in detail, an extended Crystal Ball function (CB2)~\cite{ALICE-Quarkonia-signal-extraction} is used to describe the shape of the invariant mass signal of the \jpsi and \psip. Alternatively, a pseudo-Gaussian function with a mass-dependent width is also adopted~\cite{ALICE-Quarkonia-signal-extraction}. The background continuum is empirically parameterised either with a Gaussian function with a mass dependent width (VWG) or with a fourth order polynomial times an exponential function, keeping the parameters free in the fit procedure. For $\jpsi$, the mass and width are also kept as free parameters in the fit, while the other parameters, related to the non-Gaussian tails of the mass shape, are fixed to the values obtained from Monte Carlo (MC) simulations. As a remark, the position of the mass pole of the \jpsi extracted from the fit is in excellent agreement with the PDG value~\cite{Tanabashi:2018oca} (in most cases within 1 MeV/$c^2$). As additional tests, the \jpsi tail parameters were either kept free in the fitting procedure, or fixed to those extracted from spectra corresponding to pp collisions at $\sqrt{s}=8$ TeV~\cite{Adam:2015rta}. 
 For the \psip, the mass and width are fixed to those of the $\jpsi$, since the relatively low signal to background ratio does not allow the same approach. The relations that are used are $m_{\psi(2{\rm S})} = m_{{\rm J}/\psi} + m_{\psi(2{\rm S})}^{\rm PDG} - m_{{\rm J}/\psi}^{\rm PDG}$ (where $m_{\rm i}^{\rm PDG}$ is the mass value from~\cite{Tanabashi:2018oca}) and $\sigma_{\psi(2{\rm S})} = \sigma_{{\rm J}/\psi}\cdot \sigma_{\psi(2{\rm S})}^{\rm MC}/\sigma_{{\rm J}/\psi}^{\rm MC}$, with the latter giving a 5\% increase between the \jpsi and \psip widths. This value is validated using results from a large data sample of pp collisions at $\sqrt{s}=13$ TeV~\cite{Acharya:2017hjh}, where the \psip mass and width are kept free in the fit procedure, and the observed increase between $\sigma_{{\rm J}/\psi}$ and $\sigma_{\psi(2{\rm S})}$ is also 5\%. The non-Gaussian tails used for the \jpsi are also adopted for the \psip.
 
 Various fits, combining the options described above were performed, also using two different fit ranges, in order to further test the background description (2 $< m_{\rm {\mu\mu}} <$ 5 GeV/$c^2$ and 2.2 $< m_{\rm {\mu\mu}} <$ 4.5 GeV/$c^2$).
The raw $\psip$ yields and their statistical uncertainties are taken to be the average of the results of the various performed fits, while the standard deviation of their distribution is assigned as a systematic uncertainty. An additional 5\% uncertainty, corresponding to the uncertainty on the \psip width in the large pp data sample used to validate the assumption on the relative widths for \jpsi and \psip~\cite{Acharya:2017hjh}, is quadratically added.
 
\begin{figure}[htbp]
\centering
    \includegraphics[width=70mm,height=70mm]{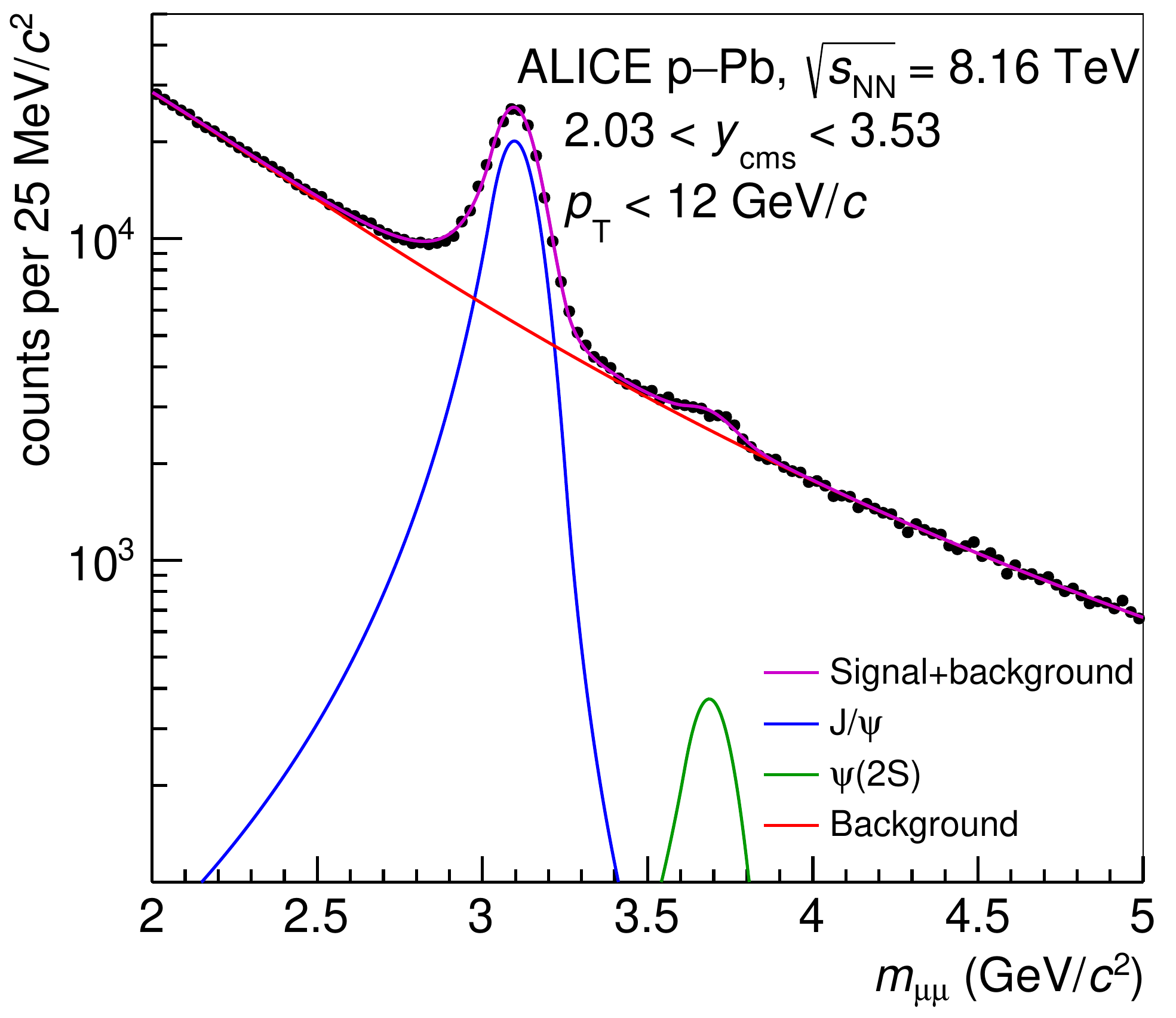}
    \qquad\qquad
    \includegraphics[width=70mm,height=70mm]{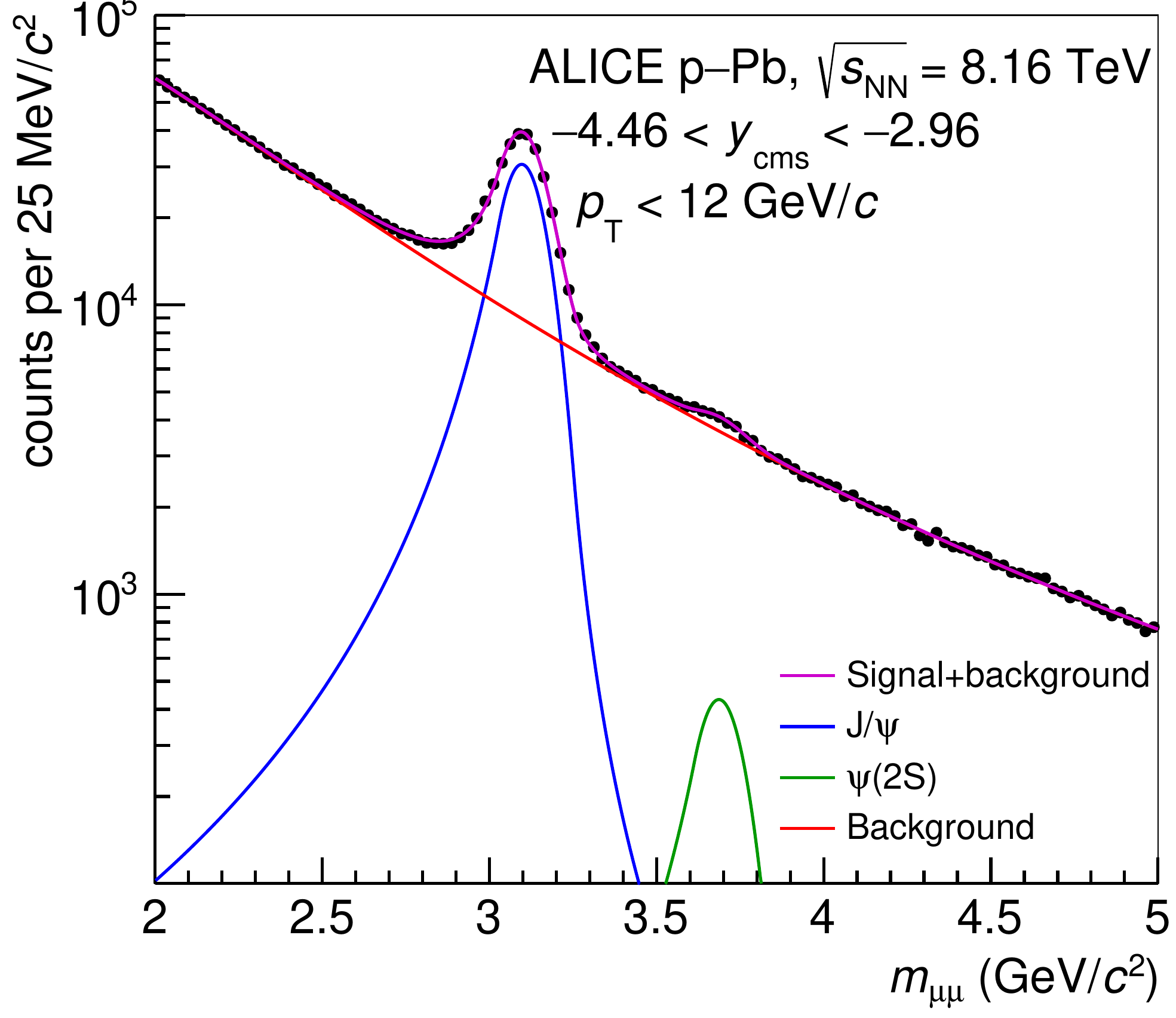}
    \caption{Fit examples of the $p_{\mathrm{T}}$ and $y$ integrated mass spectrum for the forward (left) and backward (right) rapidity data samples. The contribution of the resonances and of the background are also shown separately. These fits are performed using the CB2 as signal function and the VWG background shape.}
    \label{fig:integrated_spectra_pPbfit_Jhuma}
\end{figure}

 For the two rapidity intervals under study, the values $N^{\psi(2{\rm S})}_{\rm{pPb}} = 3148 \pm 253 \pm 243$ and $N^{\psi(2{\rm S})}_{\rm{Pbp}} = 3595 \pm 283 \pm 368$ were determined, with the first and second uncertainties being statistical and systematic. The measurement is performed in the dimuon pair transverse momentum range $p_{\rm T}<12$ GeV/$c$. As an example, Fig.~\ref{fig:integrated_spectra_pPbfit_Jhuma} shows fits to the invariant mass spectra for the two $y_{\rm cms}$ regions.
The same procedure is adopted for the evaluation of the differential yields in $y_{\rm cms}$ (2 sub-ranges each for \ppb and \pbp) and $p_{\rm T}$ (5 intervals, up to $p_{\rm T}=12$ GeV/$c$). In the interval with largest \pt  ($8<p_{\rm T}<12$ GeV/$c$) the raw $\psip$ yields are $N^{\psi(2{\rm S})}_{\rm{pPb}} = 150 \pm 39 \pm 30$ and $N^{\psi(2{\rm S})}_{\rm{Pbp}} = 131 \pm 40 \pm 33$.

 The product of acceptance and reconstruction efficiency ($A \cdot \epsilon$) for  \psip is evaluated via MC simulations, performed individually for each run, in order to correctly reproduce the evolution of the detector conditions during data taking.
 The $\pt$ and $y_{\rm cms}$ input shapes used for the simulation of \psip are tuned directly on data, by performing a differential analysis in narrower intervals and using an iterative method~\cite{Acharya:2018kxc}. The procedure is found to converge after only two iterations. The decay products of the \psip  are then propagated through a realistic description of the ALICE set-up, based on GEANT3.21~\cite{Brun:1082634}. The $A \cdot \epsilon$ values, averaged over the data taking periods and integrated over $y_{\rm cms}$ and $p_{\rm T}$, amount to 0.272 for  \ppb and 0.258 for \pbp collisions, with a negligible statistical uncertainty. The systematic uncertainties on the acceptance are evaluated by performing an alternative simulation based on the corresponding input shapes for the \jpsi~\cite{Abelev:2014zpa}. A 3\% and 1.5\% effect is found for \ppb and \pbp, respectively. When considering differential values as a function of $y_{\rm cms}$ and $p_{\rm T}$, the uncertainties vary between 0.4--4.0\% (0.1--4.4\%) for \ppb (\pbp). The reconstruction efficiency is the product of trigger, tracking and matching efficiency terms. The latter term refers to the procedure used to pair tracks reconstructed in the tracking system with the corresponding track segments in the trigger detector.
 The systematic uncertainties on the three efficiencies mentioned above are evaluated in the same way, and have the same values as those reported for the \jpsi analysis~\cite{Acharya:2018kxc}. The largest contribution is that from the trigger which amounts to 2.6\% (3.1\%) for the integrated \ppb (\pbp) data sample.

The integrated luminosities for the two data samples, as detailed in Ref.~\cite{Acharya:2018kxc}, are obtained from $L_{\rm int} = N_{\rm MB}/\sigma_{\rm MB}$, where $N_{\rm MB}$ is the number of MB events and $\sigma_{\rm MB}$ the cross section corresponding to the MB trigger condition, obtained through a van der Meer scan~\cite{ALICE-PUBLIC-2018-002}. The $N_{\rm MB}$ quantity was estimated as the number of analysed dimuon triggers times the inverse of the probability of having a triggered dimuon in a MB event. These  values are quoted in Ref.~\cite{Acharya:2018kxc}.

The suppression of \psip with respect to the corresponding pp yield is quantified by the nuclear modification factor $R_{\rm{pPb}}^{\psi(2{\rm S})}$. Its evaluation is performed through the following expression:
\begin{equation} \label{eqn:raa}
R_{\rm{pPb}}^{\psi(2{\rm S})}(p_{\rm{T}},y_{\rm{cms}}) = \frac{{\rm d^{2}}\sigma_{\rm{pPb}}^{\psi(2{\rm S})}/{\rm d} p_{\rm{T}}{\rm d} y_{\rm{cms}}}{{\rm A}_{\rm{Pb}}\cdot{\rm  d}^{2}\sigma_{\rm{pp}}^{\psi(2{\rm S})}/{\rm d}p_{\rm{T}}{\rm d}y_{\rm{cms}}}
\end{equation}
where $A_{\rm{Pb}}=208$ is the mass number of the lead nucleus and the production cross sections in \ppb and pp are evaluated at the same collision energy and in the same kinematic domain. For this analysis, the \psip production cross section in pp collisions, integrated over \pt and for each of the two rapidity ranges is evaluated from the average of the \jpsi cross sections measured by ALICE~\cite{Adam:2015rta} and LHCb~\cite{Aaij:2013yaa} at $\sqrt{s}= 8$ TeV, multiplied by the ratio of cross sections $[\sigma_{\psi(2{\rm S})}/\sigma_{{\rm J}/\psi}]_{\rm{pp}}$, obtained via an interpolation of ALICE results at $\sqrt{s}=$ 5, 7, 8 and 13 TeV~\cite{Acharya:2017hjh} assuming no energy dependence. The interpolation is in very good agreement with the pp results, and allows the uncertainties on this quantity to be significantly reduced. To account for the slight difference in collision energy between pp and \ppb data (8 TeV vs 8.16 TeV) a 1.5\% correction factor on the \jpsi cross section at $\sqrt{s}=8$ TeV is introduced, obtained from an interpolation of \jpsi production cross sections measured at various $\sqrt{s}$~\cite{Acharya:2017hjh}. Finally, both the \jpsi cross section and the $[\sigma^{\psi(2{\rm S})}/\sigma^{{\rm J}/\psi}]_{\rm{pp}}$ ratio must be evaluated in the rapidity domain covered by the \ppb and \pbp configurations. For the \jpsi cross section, the procedure detailed in Ref.~\cite{Acharya:2018kxc} and based on a polynomial or Gaussian interpolation of the $y_{\rm cms}$-dependence is adopted. For the ratio $[\sigma^{\psi(2{\rm S})}/\sigma^{{\rm J}/\psi}]_{\rm{pp}}$ a small correction factor, related to the slightly different rapidity distributions for \jpsi and \psip, as discussed in Ref.~\cite{Abelev:2014zpa}, and amounting to $\sim 1$\%, is assigned as a systematic uncertainty. Other systematic uncertainties related to  $[\sigma^{\psi(2{\rm S})}/\sigma^{{\rm J}/\psi}]_{\rm{pp}}$ include a term (6.0\%) corresponding to the uncertainty on the interpolation procedure and a further 1\% obtained by assuming, rather than a flat $\sqrt{s}$ dependence of the ratio, the one calculated by NRQCD+CGC models~\cite{Ma:2010yw,Ma:2014mri} as quoted in Ref.~\cite{Acharya:2017hjh}. Finally, there is a contribution from the uncertainty on the \jpsi cross section in pp collisions at $\sqrt{s}=8$ TeV~(7.3\% for both \ppb and \pbp, see Table~1 of ~\cite{Acharya:2018kxc}).

The evaluation of the reference cross section in the rapidity sub-intervals and as a function of $p_{\rm T}$ is performed with the same procedure summarised above. More in detail, for each $y_{\rm cms}$ and $p_{\rm T}$ interval, pp results at various $\sqrt{s}$ are again interpolated with a constant function, which is found to well reproduce the data. For this differential study, the relatively small data sample for pp collisions at $\sqrt{s}=5.02$ TeV~\cite{Acharya:2017hjh} is not used in the interpolation.

A summary of the systematic uncertainties on the determination of the \psip cross sections and of the nuclear modification factor is given in 
Tab.~\ref{table:systematics}. The contribution from the signal extraction procedure is the largest, and is uncorrelated among the various $p_{\rm T}$ and $y_{\rm cms}$ intervals. The uncertainties on the MC input shapes and on the various efficiencies are also considered as uncorrelated as a function of $p_{\rm T}$ and $y_{\rm cms}$. The uncertainties on the \ppb luminosity values correspond to those quoted in Ref.~\cite{Acharya:2018kxc}. Concerning the pp reference, the uncertainties corresponding to the luminosity measurement affecting the \jpsi cross sections in pp are correlated~\cite{Acharya:2018kxc}, while the remaining contributions are uncorrelated  over $y_{\rm cms}$ and $p_{\rm T}$. The various uncorrelated and correlated uncertainties are added in quadrature and separately quoted in the numerical results and in the figures of the next section.

\begin{table}[!h]
\begin{center}
\caption{Systematic uncertainties on the  determination of the \psip cross sections times branching ratio and nuclear modification factors, shown separately for the \ppb and \pbp configurations. When a single value is quoted, it refers to quantities that have no \pt or $y_{\rm cms}$ dependence. In the other cases, the number outside parentheses is for integrated quantities, while the ranges in parentheses indicate the variation of the systematic uncertainties in the \pt and $y_{\rm cms}$ intervals.}
\begin{tabular}{|c|c|c|}
\hline
source & p--Pb (\%) & Pb--p (\%)\\ \hline
signal extraction & 7.7 (8.0--20.0) & 10.2 (9.1--24.9) \\ \hline
trigger efficiency& 2.6 (1.0--5.0) & 3.1 (1.0--6.0)\\ \hline
tracking efficiency& 1.0 & 2.0 \\ \hline
matching efficiency& 1.0 & 1.0 \\ \hline
MC input & 3 (0.4--4.0)& 1.5 (0.1--4.4)\\ \hline
$L_{\rm int}^{\rm pPb}$ {\rm (corr.)} & 0.5 & 0.7 \\ \hline
$L_{\rm int}^{\rm pPb}$ {\rm (uncorr.)} & 2.1 & 2.2 \\ \hline
pp reference (corr.) & 7.1 & 7.1 \\ \hline
pp reference (uncorr.) & 6.3 (7.0--11.8) & 6.5 (7.2--11.9) \\ \hline
\end{tabular}
\label{table:systematics}
\end{center}
\end{table}

\section{Results}\label{section:results}

\begin{figure}[htbp]
    \centering
    \includegraphics[width=0.6\linewidth]{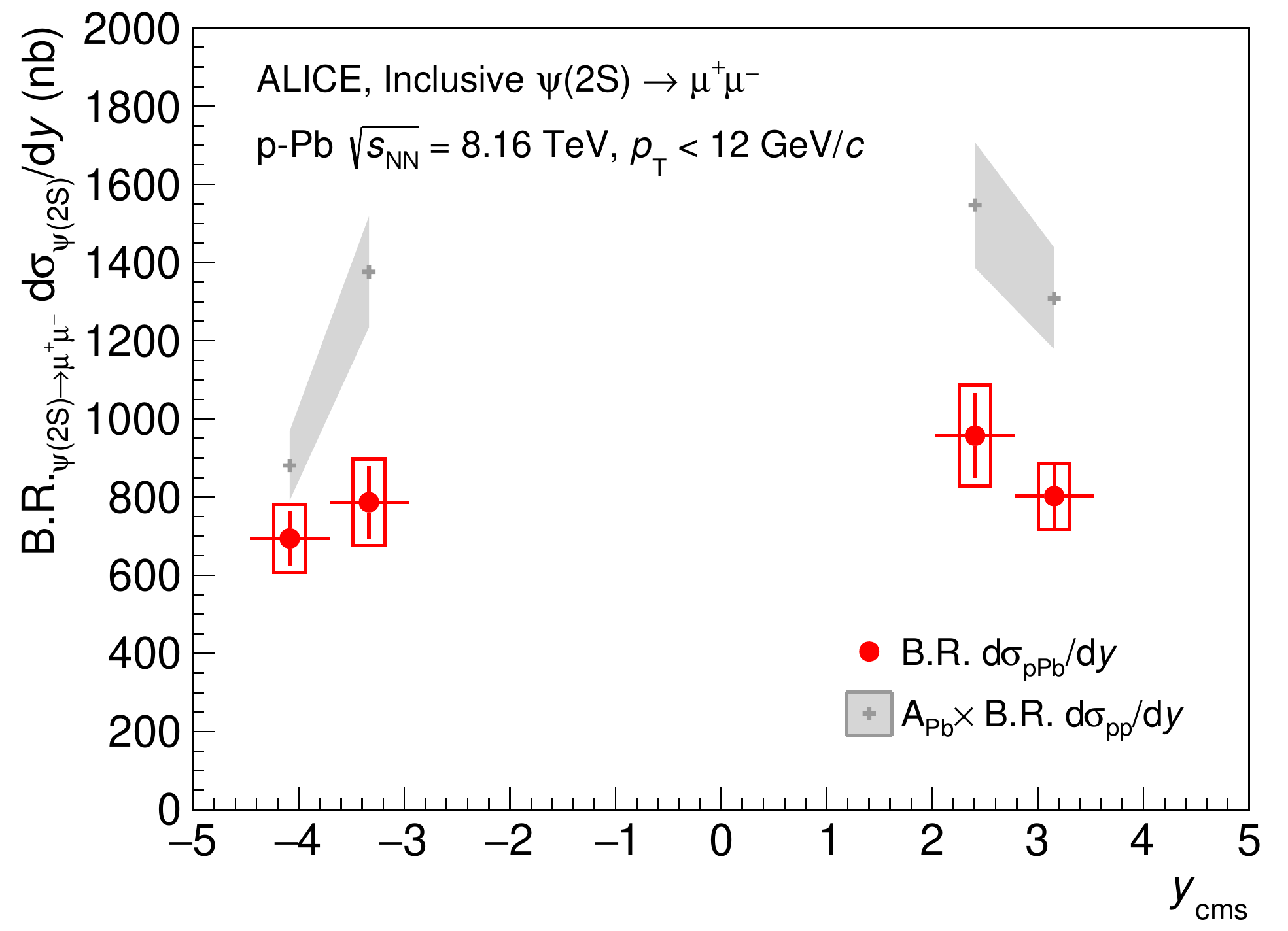}

       \caption{The differential cross section times branching ratio ${\rm B.R.}_{\psi(2{\rm S})\rightarrow\mu^+\mu^-}{\rm d}\sigma^{\psi(2{\rm S})}/{\rm d}y$ for $p_{\rm T}<12$ GeV/$c$. The error bars represent the statistical uncertainties, while the boxes correspond to total  systematic uncertainties. The latter are uncorrelated among the points, except for a very small correlated uncertainty (0.5\% and 0.7\% for the forward and backward $y_{\rm cms}$ samples, respectively). The grey bands correspond to the reference pp cross section scaled by A$_{\rm Pb}$. } 
    \label{fig:diffcrosssecty}
\end{figure}

\begin{figure}[htbp]
    \centering
    \includegraphics[width=0.5\linewidth]{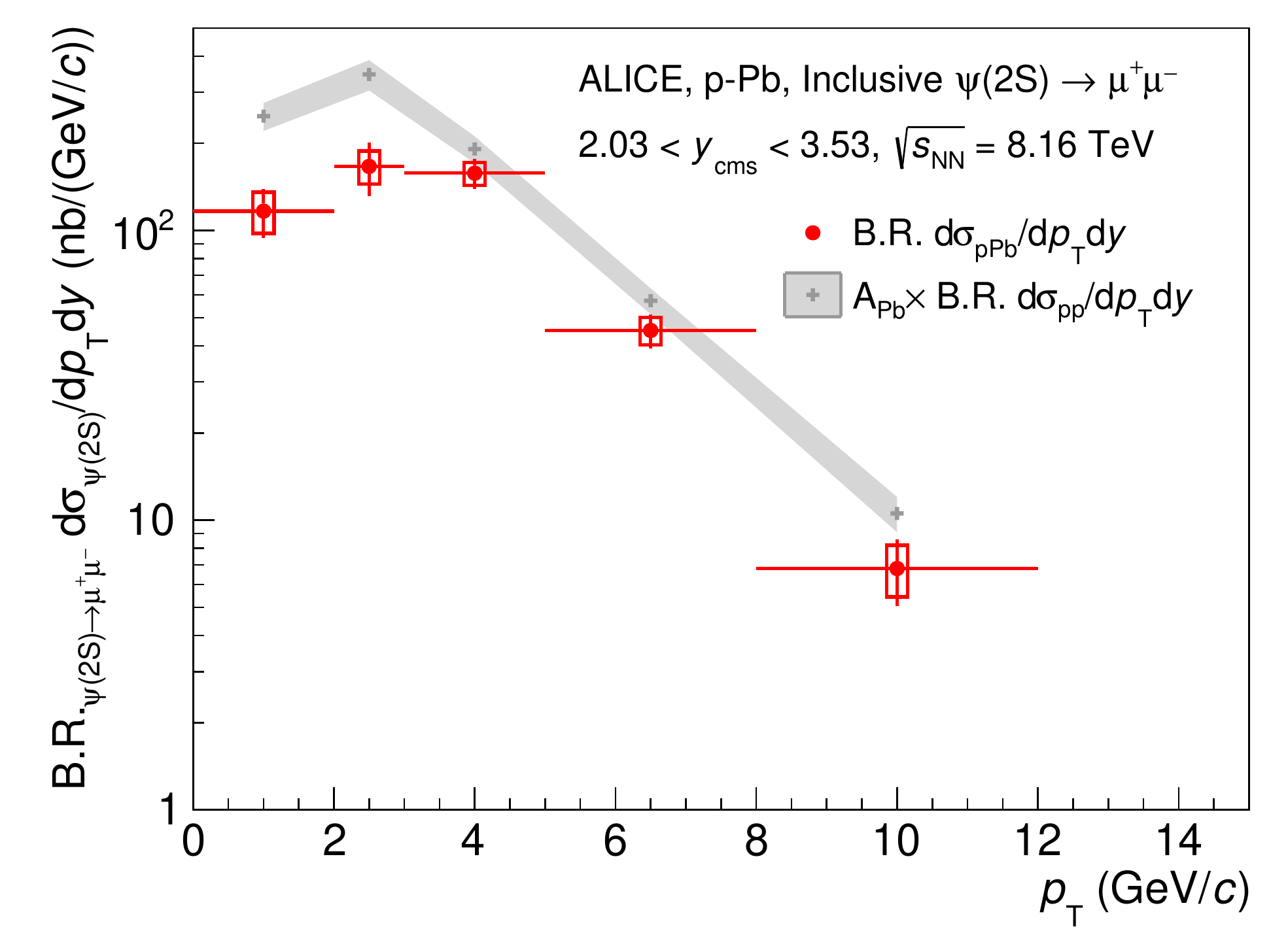}
    \includegraphics[width=0.5\linewidth]{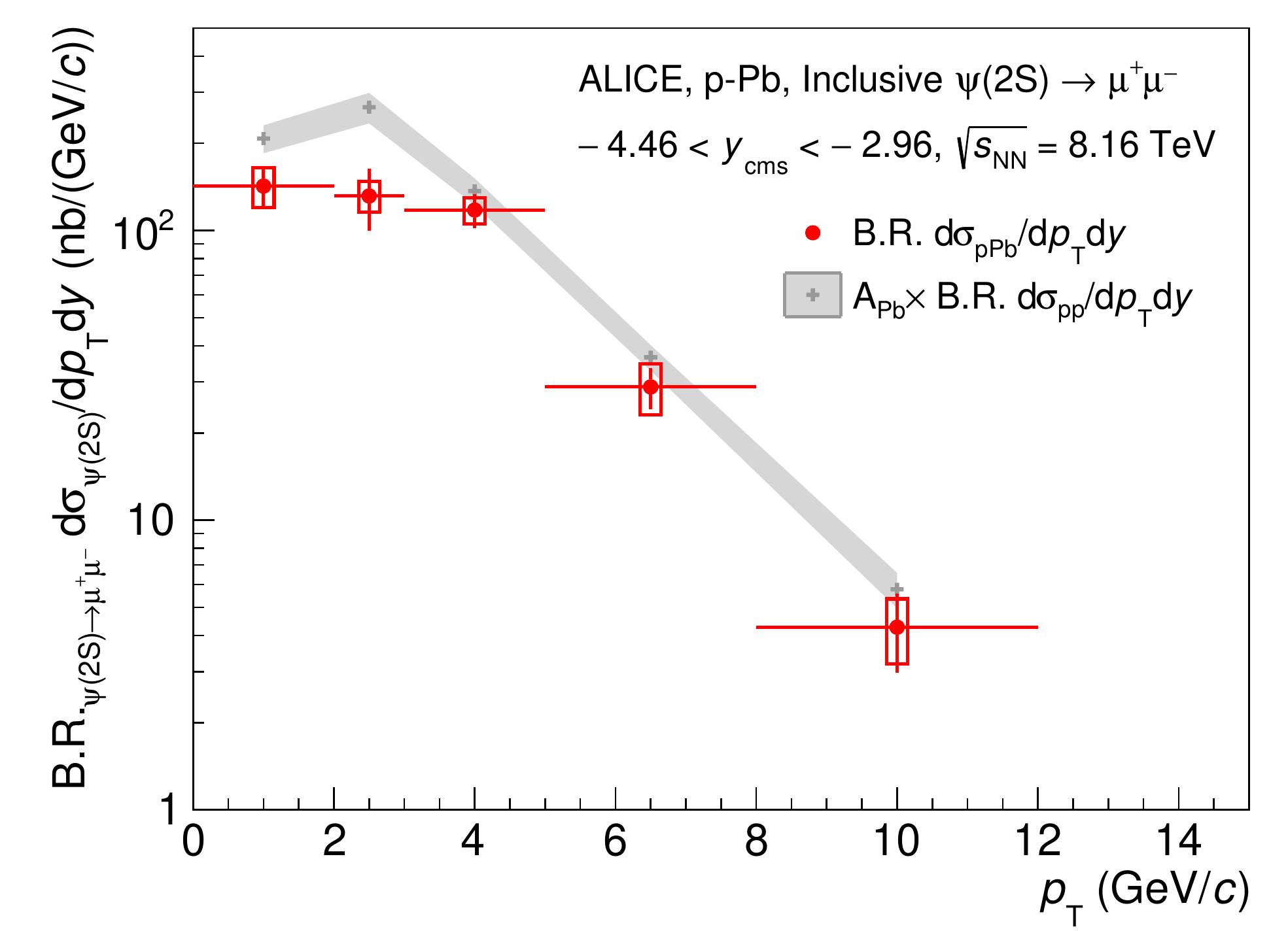}

       \caption{The differential cross sections ${\rm B.R.}_{\psi(2{\rm S})\rightarrow\mu^+\mu^-}{\rm d}^2\sigma^{\psi(2{\rm S})}/{\rm d}y{\rm d}p_{\rm T}$ for \ppb collisions at $\sqrtSnn = 8.16$ TeV, shown separately for the forward and backward $y_{\rm cms}$ samples. The error bars represent the statistical uncertainties, while the boxes correspond to total systematic uncertainties. The latter are uncorrelated among the points, except for a very small correlated uncertainty (0.5\% and 0.7\% for the forward and backward $y_{\rm cms}$ samples, respectively). The grey bands correspond to the reference pp cross section scaled by A$_{\rm Pb}$.} 
    \label{fig:diffcrosssectpt}
\end{figure}

The measured inclusive \psip production cross sections for \ppb collisions at $\sqrt{s_{\rm NN}}=8.16$ TeV, multiplied by the branching ratio to muon pairs and integrated over $p_{\rm T}<12$ GeV/$c$ are:
\begin{eqnarray*}
{\rm B.R.}_{\psi(2{\rm S})\rightarrow\mu^+\mu^-}\cdot\sigma^{\psi(2{\rm S})}_{\rm{pPb}} (2.03 < y_{\rm{cms}} < 3.53) = 1.337 \pm 0.108 \pm 0.121 \pm 0.007\, \mu{\rm b}\\
{\rm B.R.}_{\psi(2{\rm S})\rightarrow\mu^+\mu^-}\cdot\sigma^{\psi(2{\rm S})}_{\rm{Pbp}} (-4.46 < y_{\rm{cms}} < -2.96) = 1.124 \pm 0.089 \pm 0.126 \pm 0.008\, \mu{\rm b}
\end{eqnarray*}

where the first uncertainty is statistical, the second and third are uncorrelated and correlated systematic, respectively.
The differential \psip cross sections are determined as a function of $y_{\rm cms}$ (splitting the forward and backward intervals in two sub-intervals) and $p_{\rm T}$ (5 intervals). The results are shown in Figs.~\ref{fig:diffcrosssecty} and~\ref{fig:diffcrosssectpt}.
The reported values include, in addition to the prompt component, a contribution from the decays of b-hadrons, which was shown by LHCb in \ppb collisions at $\sqrt{s_{\rm NN}}=5.02$ TeV~\cite{Aaij:2016eyl} to amount to $\sim$20--30\% of the inclusive cross section.
Furthermore, Figs.~\ref{fig:diffcrosssecty} and~\ref{fig:diffcrosssectpt} also show, as a band, the reference pp cross section obtained with the interpolation procedure described in the previous section, scaled by A$_{\rm Pb}$.


\begin{figure}[!h]
    \centering
    \includegraphics[width=0.6\linewidth]{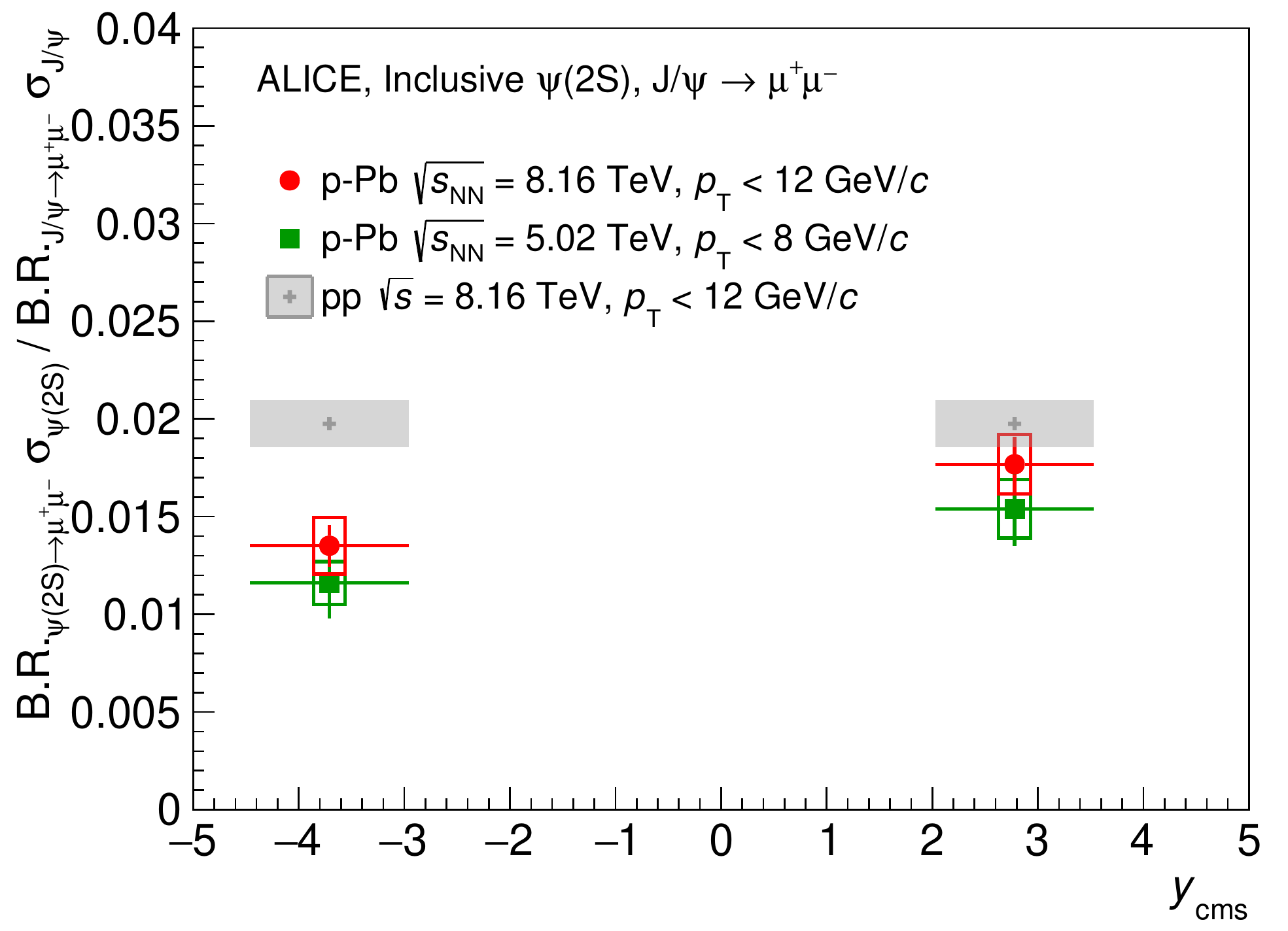}
       \caption{The ratio ${\rm B.R.}_{\psi(2{\rm S})\rightarrow\mu^+\mu^-}\sigma^{\psi(2{\rm S})}/{\rm B.R.}_{{\rm J}/\psi\rightarrow\mu^+\mu^-}\sigma^{{\rm J}/\psi}$ as a function of  $y_{\rm cms}$ for \ppb collisions at $\sqrtSnn = 8.16$ TeV, compared with the corresponding pp quantity, shown as a grey band and obtained via an interpolation of results at $\sqrt{s}=$ 5, 7, 8 and 13 TeV~\cite{Acharya:2017hjh}. The error bars represent the statistical uncertainties, while the boxes correspond to uncorrelated systematic uncertainties. The published \ppb results at $\sqrtSnn  = 5.02$ TeV~\cite{Abelev:2014zpa} are also shown.}
    \label{fig:integrated ratio}
\end{figure}

The ratio of the \psip and \jpsi cross sections is an interesting quantity for the comparison of the production of the two resonances across different systems, because the terms related to the luminosity and efficiencies and the corresponding uncertainties cancel. It has been computed in this analysis  as the ratio of the acceptance-corrected number of \psip and \jpsi. In Fig. \ref{fig:integrated ratio} the $p_{\rm T}$-integrated cross section ratio is shown for the two rapidity intervals. In the same figure, this quantity is compared with the corresponding pp result at the same collision energy, obtained through the interpolation procedure described in the previous section. At backward rapidity, the ratio is significantly lower (2.9$\sigma$ effect) than in pp, while at forward rapidity the values are compatible. In the same figure, the results are compared with those obtained in \ppb collisions at $\sqrt{s_{\rm NN}}=5.02$ TeV~\cite{Abelev:2014zpa}. No $\sqrt{s_{\rm NN}}$-dependence can be observed within uncertainties.

\begin{figure}[!h]
    \centering
    \includegraphics[width=0.5\linewidth]{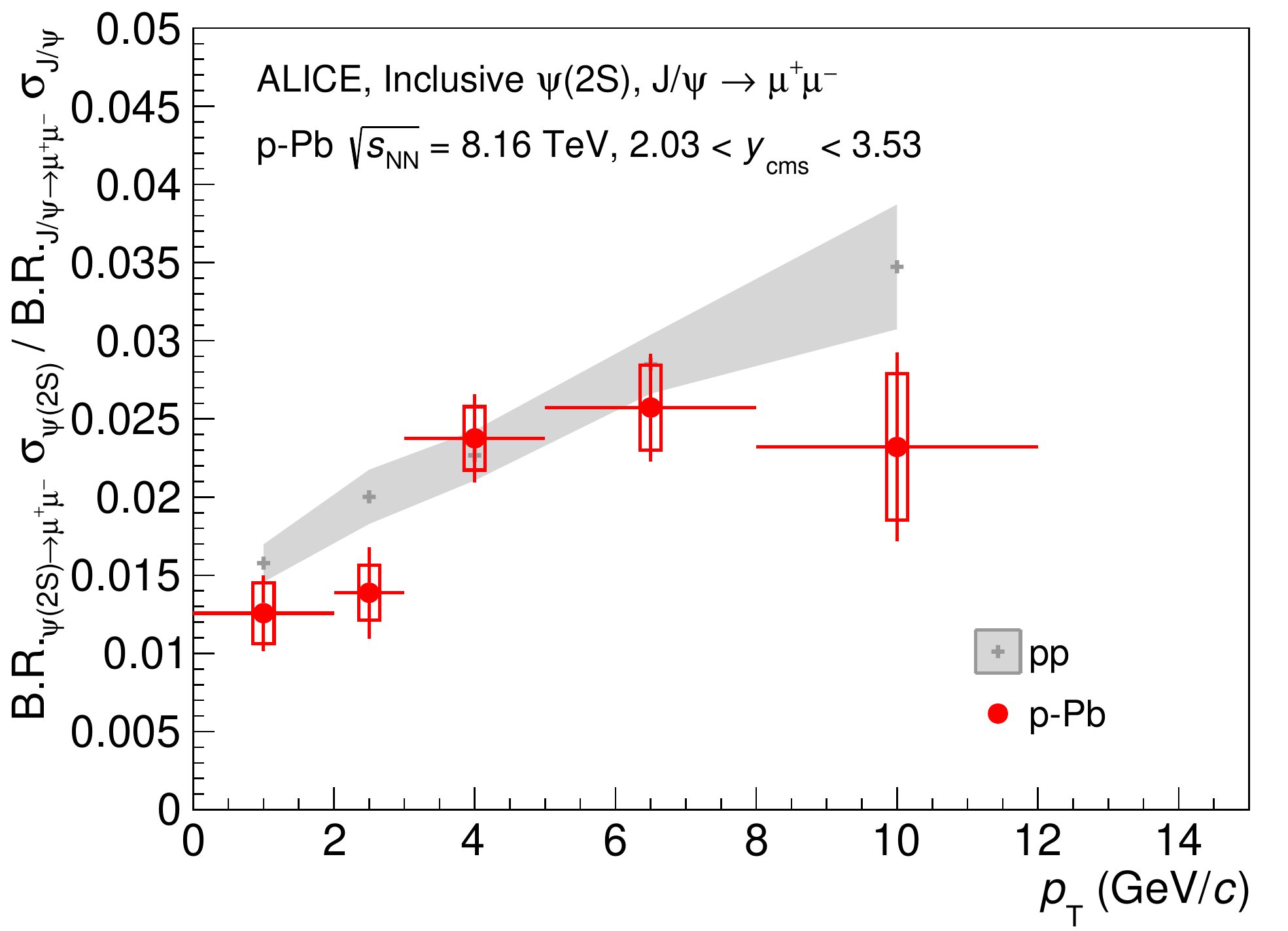}
    \includegraphics[width=0.5\linewidth]{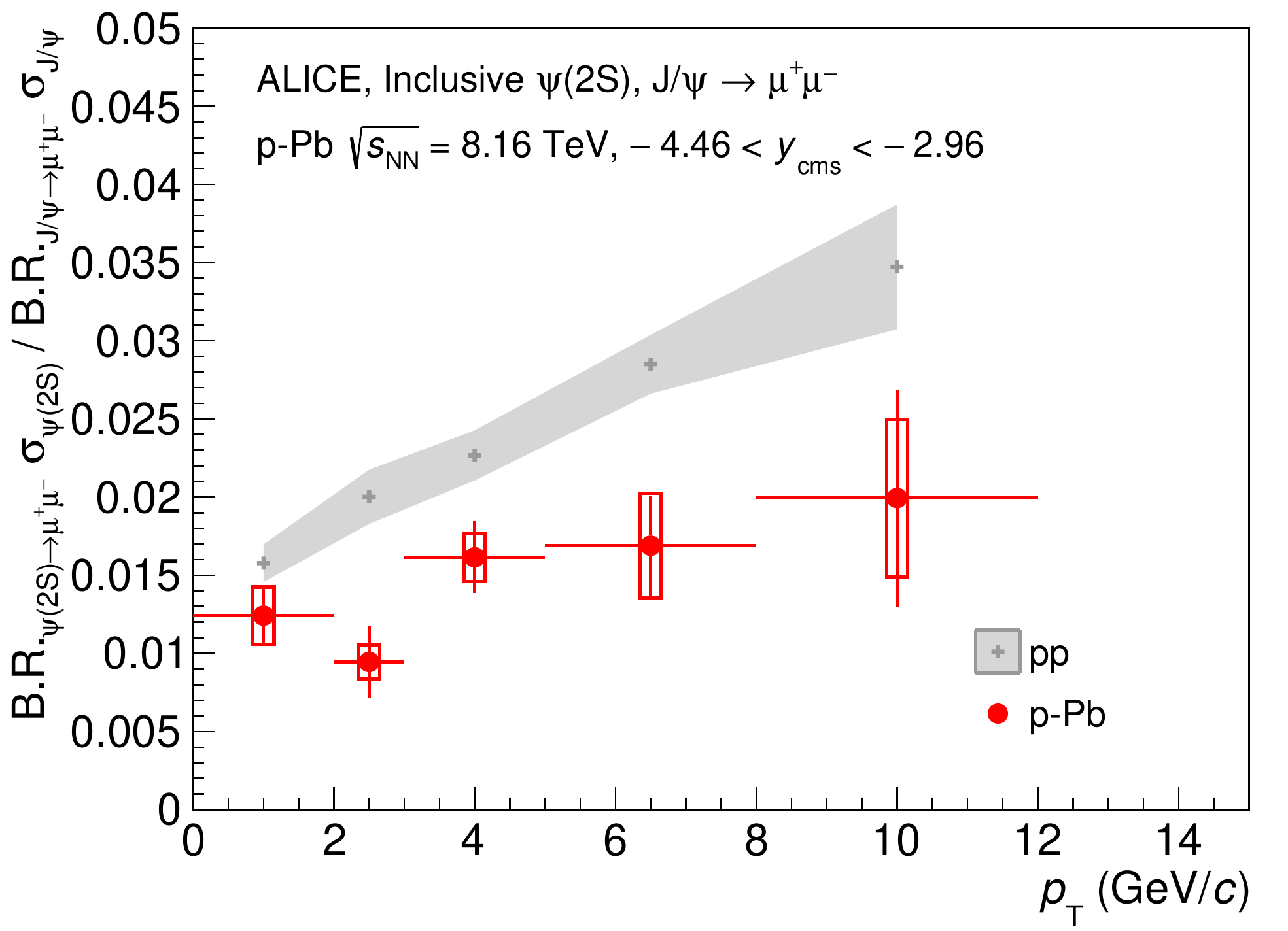}
      \caption{The ratio ${\rm B.R.}_{\psi(2{\rm S})\rightarrow\mu^+\mu^-}\sigma^{\psi(2{\rm S})}/{\rm B.R.}_{{\rm J}/\psi\rightarrow\mu^+\mu^-}\sigma^{{\rm J}/\psi}$ as a function of \pt, for \ppb collisions at $\sqrtSnn = 8.16$ TeV, compared with the corresponding pp quantity, shown as a grey band and obtained via an interpolation of results at $\sqrt{s}=$ 7, 8 and 13 TeV~\cite{Acharya:2017hjh}. The error bars represent the statistical uncertainties, while the boxes correspond to uncorrelated systematic uncertainties.}
    \label{fig:pTdependent ratio}
\end{figure}

In Fig.~\ref{fig:pTdependent ratio} the $p_{\rm T}$-dependence of the ratio of the \psip and \jpsi cross section is shown. It is compared with the corresponding pp ratio obtained through the interpolation procedure described in the previous section. Also here a stronger relative suppression of \psip with respect to \jpsi is visible at backward rapidity.

The suppression of $\psip$ can be more directly quantified by considering  the nuclear modification factors, estimated following the procedure described in the previous section. 
The numerical values, integrated over the interval $p_{\rm T}<12$ GeV/$c$, are:
\begin{eqnarray*}
R^{\psi(2{\rm S})}_{\rm{pPb}} (2.03 < y_{\rm{cms}} < 3.53) = 0.628\pm 0.050 \,{\rm (stat.)} \pm 0.069 \,{\rm (syst. uncorr.)} \pm 0.045 \,{\rm (syst. corr.)}\\
R^{\psi(2{\rm S})}_{\rm{Pbp}} (-4.46 < y_{\rm{cms}} < -2.96) = 0.684\pm 0.054 \,{\rm (stat.)} \pm 0.088 \,{\rm (syst. uncorr.)} \pm 0.049 \,{\rm (syst. corr.)}
\end{eqnarray*}

\begin{figure}[htbp]
\begin{center}     
\includegraphics[width=0.49\linewidth]{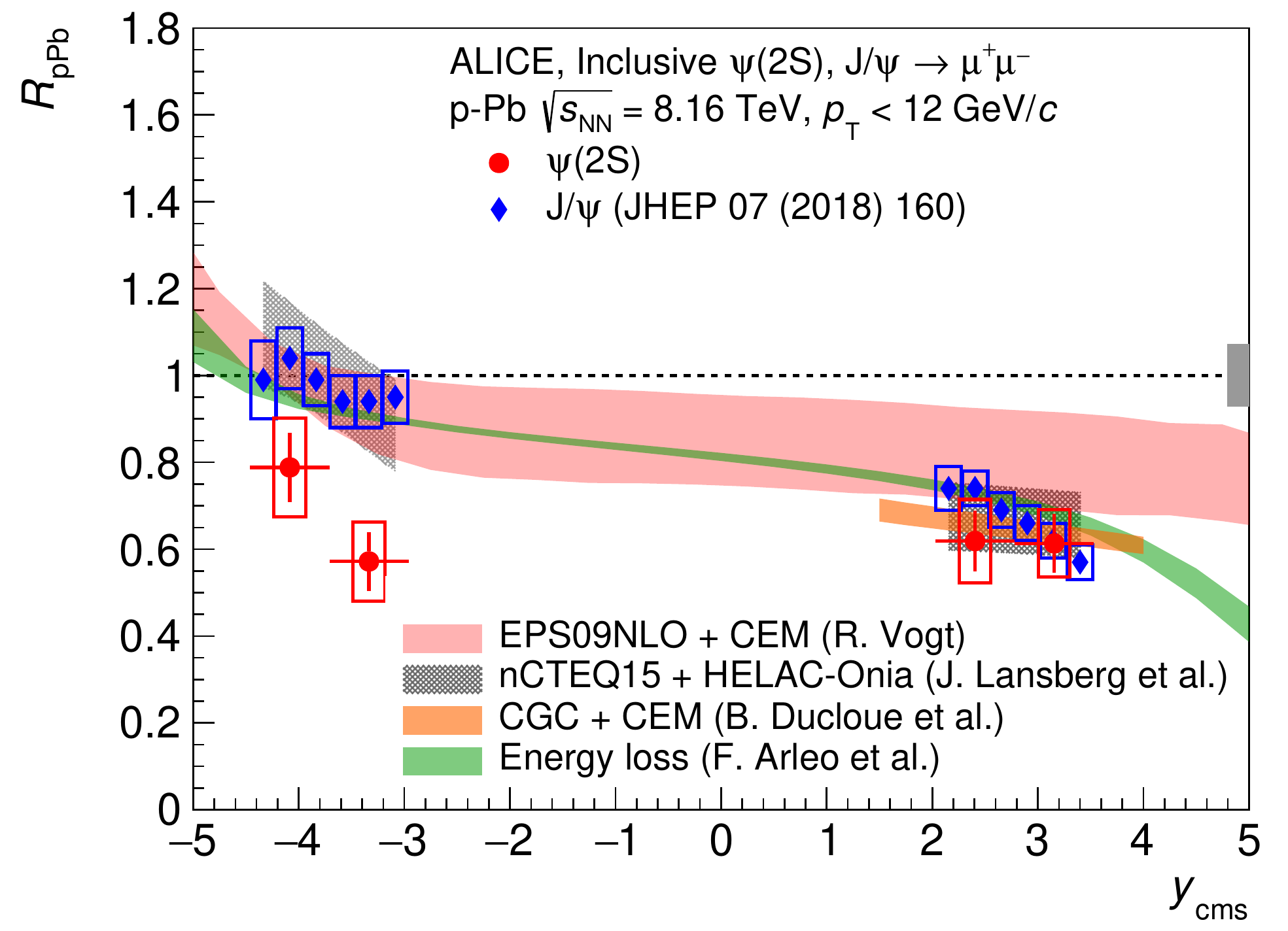}
\includegraphics[width=0.49\linewidth]{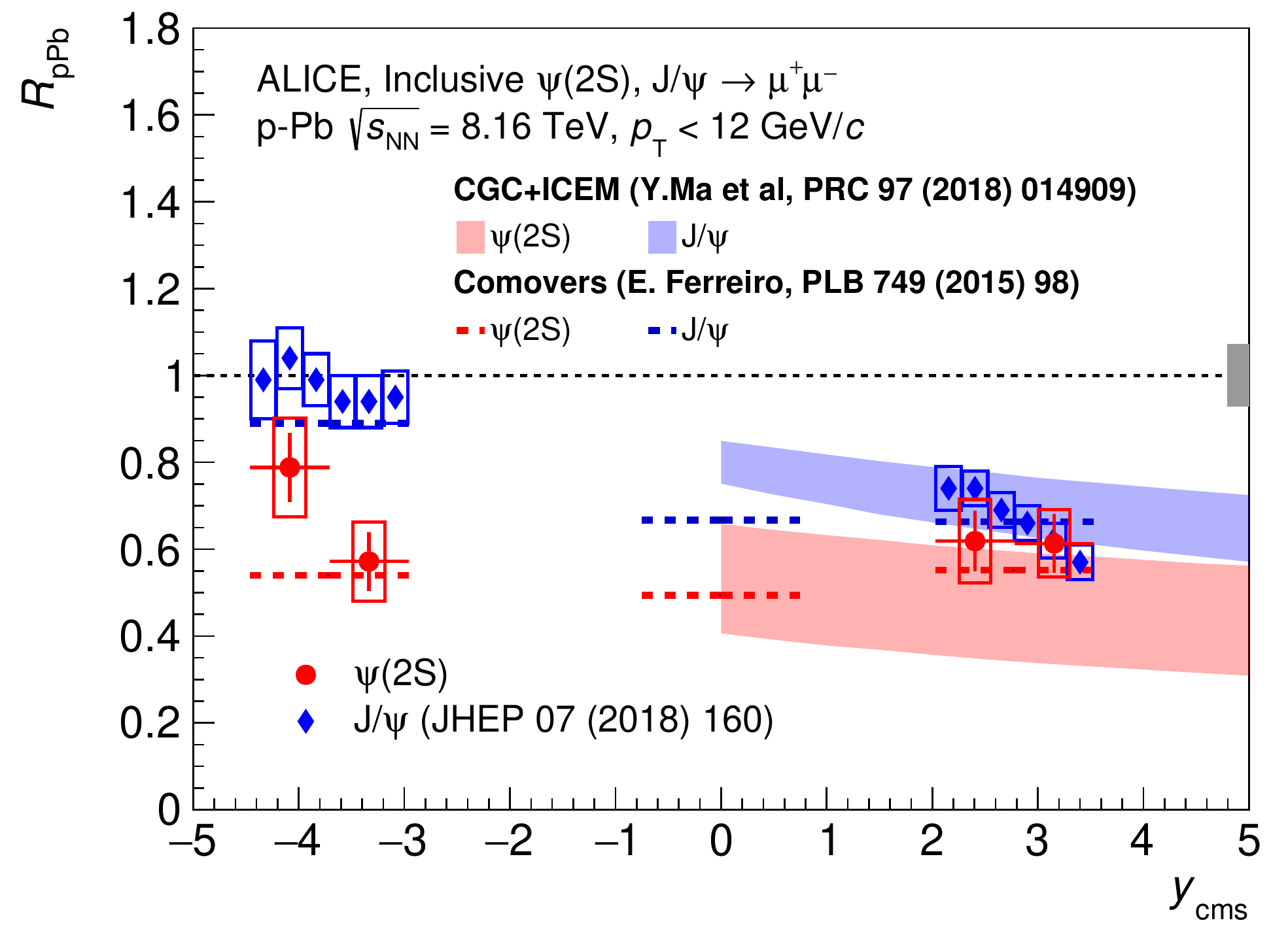}
\caption{The $y_{\rm cms}$-dependence of $R_{\rm{pPb}}$ for \psip and \jpsi~\cite{Acharya:2018kxc} in p--Pb collisions at $\sqrtSnn  = 8.16$ TeV. The error bars represent the statistical uncertainties, while the boxes correspond to uncorrelated systematic uncertainties  and the box at $R_{\rm pPb}=1$ to correlated systematic uncertainties. The results are compared with models including initial-state effects~\cite{Ducloue:2016pqr,Albacete:2017qng,Kusina:2017gkz} and coherent energy loss~\cite{Arleo:2014oha,Albacete:2017qng} (left panel), and to models which also implement final-state effects~\cite{Ma:2017rsu,Ferreiro:2014bia} (right panel). }
\label{fig:RAA_y_jpsi_CNM}
\end{center}
\end{figure}

The reported values refer to inclusive production. It was shown by LHCb, when studying \ppb collisions at $\sqrt{s_{\rm NN}}=5.02$ TeV, that inclusive and prompt nuclear modification factors are compatible within uncertainties~\cite{Aaij:2016eyl}.
In Fig.~\ref{fig:RAA_y_jpsi_CNM},  $R_{\rm{pPb}}^{\psi(2{\rm S})}$ is shown splitting the forward and backward rapidity samples in two intervals. The results are compared with those for $R_{\rm{pPb}}^{{\rm J}/\psi}$~\cite{Acharya:2018kxc}. 
For $\psip$, the suppression reaches up to 30--40\% and is compatible, within uncertainties, at forward and backward $y_{\rm cms}$. Relatively to J/$\psi$, a stronger suppression is visible at backward rapidity, whereas the results are compatible at forward rapidity. The data are also compared (left panel) with theoretical calculations based on initial-state effects or coherent energy loss, whose output is largely independent on the specific charmonium resonance, and can therefore be compared with both \jpsi and \psip results.  Calculations based on the CGC approach~\cite{Ducloue:2016pqr,Albacete:2017qng}, on nuclear shadowing~\cite{Albacete:2017qng,Kusina:2017gkz}, implemented according to different parameterizations (EPS09NLO~\cite{Eskola:2009uj}, nCTEQ15~\cite{Kovarik:2015cma}) or finally on coherent energy loss~\cite{Arleo:2014oha,Albacete:2017qng}, show good agreement with the \jpsi results but fail to describe the \psip $R_{\rm pPb}$ at backward rapidity.

The possible influence of final-state interactions, leading to a break-up of the charmonium resonances, is taken into account in theory calculations where these effects are due to either soft color exchanges in the hadronizing c$\overline{\rm{c}}$ pair~\cite{Ma:2017rsu}, or final-state interactions with the comoving medium~\cite{Ferreiro:2014bia}. The former calculation describes the initial state in terms of a CGC state, and results are available only at forward rapidity, corresponding to low Bjorken-$x$ values in the Pb nucleus, where the system may be described using this approach. The two models reach a fair agreement with data for both $\psip$ and J/$\psi$, as shown in the right panel of  Fig.~\ref{fig:RAA_y_jpsi_CNM}.

\begin{figure}[htbp]
    \centering
    \includegraphics[width=0.5\linewidth]{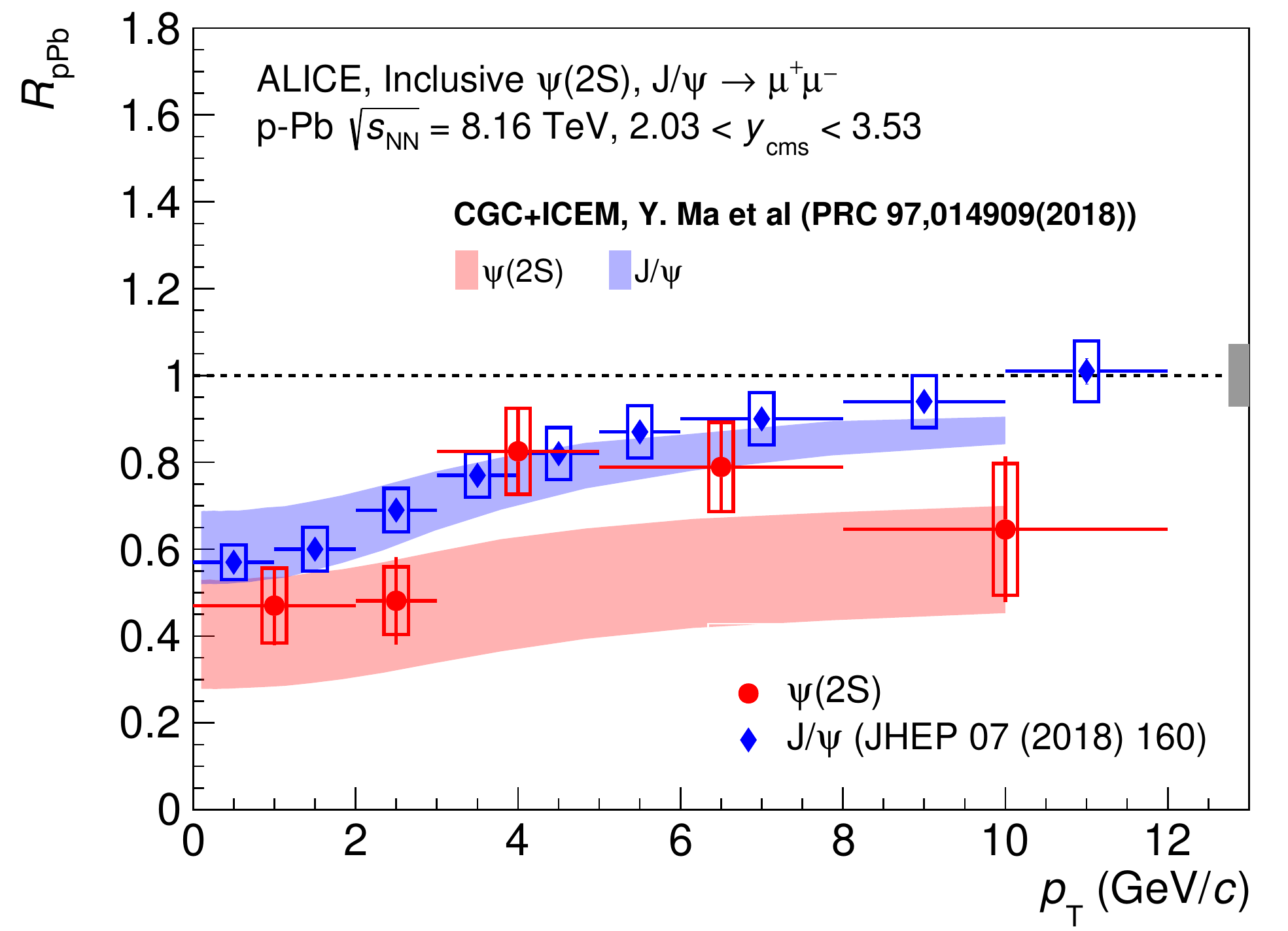}
    \includegraphics[width=0.5\linewidth]{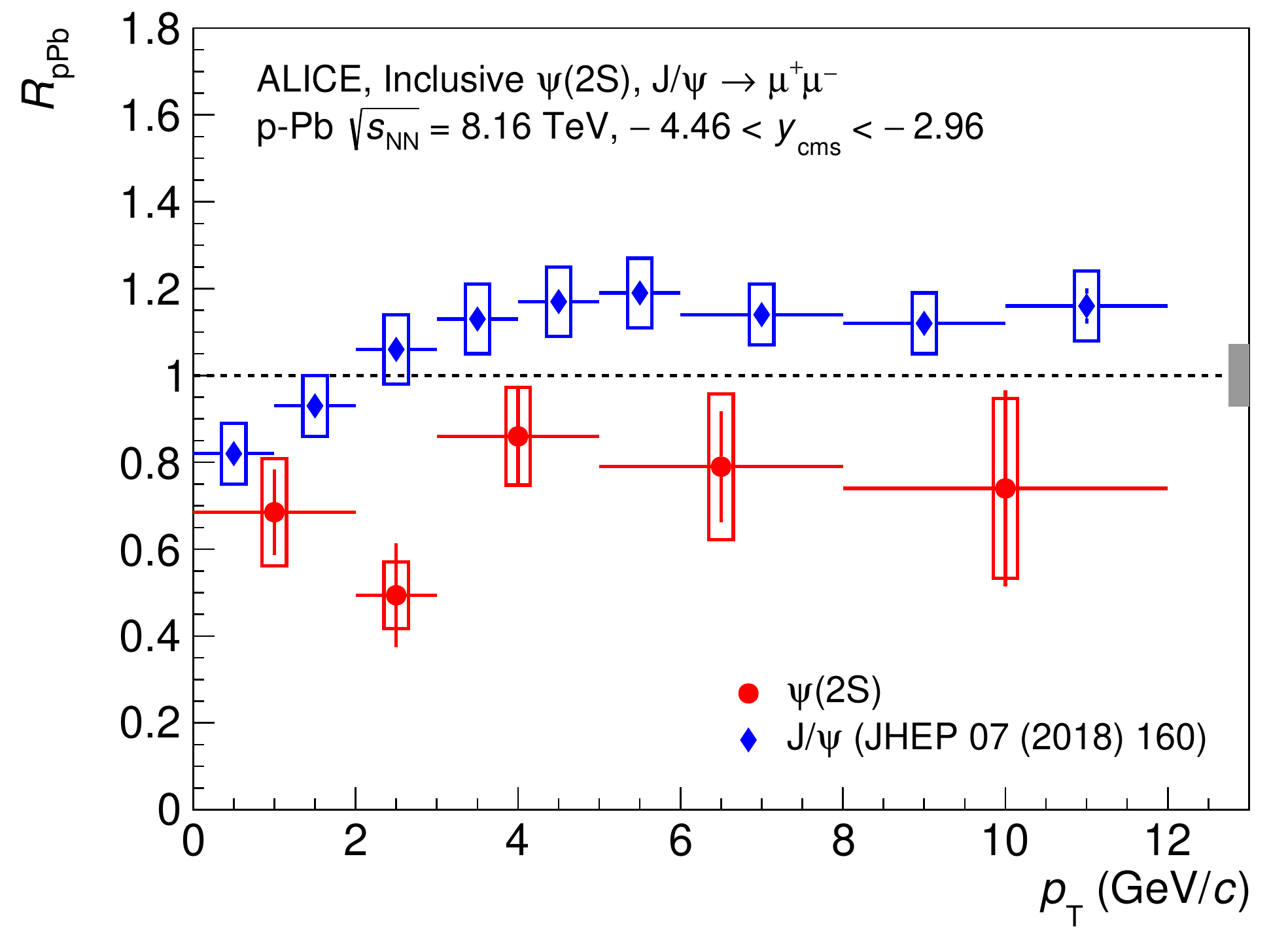}
       \caption{The $p_{\rm T}$-dependence of $R_{\rm{pPb}}$ for \psip and \jpsi at forward (left) and backward (right) rapidity in \ppb collisions, at $\sqrtSnn  = 8.16$ TeV. The error bars represent the statistical uncertainties, while the boxes correspond to uncorrelated systematic uncertainties  and the box at $R_{\rm pPb}=1$ to correlated systematic uncertainties. The comparison with the results of a CGC-based model~\cite{Ma:2017rsu}, which implements final-state effects, is also shown.}
    \label{fig:RpPb_pt_jpsi}
\end{figure}

The present data sample allows a \pt-differential study of $R^{\psi(2{\rm S})}_{\rm{pPb}}$ up to $p_{\rm T} = 12$ GeV/$c$. The results are plotted in Fig.~\ref{fig:RpPb_pt_jpsi}, separately for forward and backward rapidity, and compared with published results for J/$\psi$~\cite{Acharya:2018kxc}. At forward rapidity the $\psip$ suppression is compatible with that of J/$\psi$,
while at backward rapidity the $\psip$ suppression, which is independent of \pt within uncertainties, is significantly stronger.
The CGC-based model~\cite{Ma:2017rsu} results are found to fairly  match the experimental findings. No theory comparison is yet available for backward rapidity.

\begin{figure}[htbp]
\begin{center}     
\includegraphics[width=0.6\linewidth]{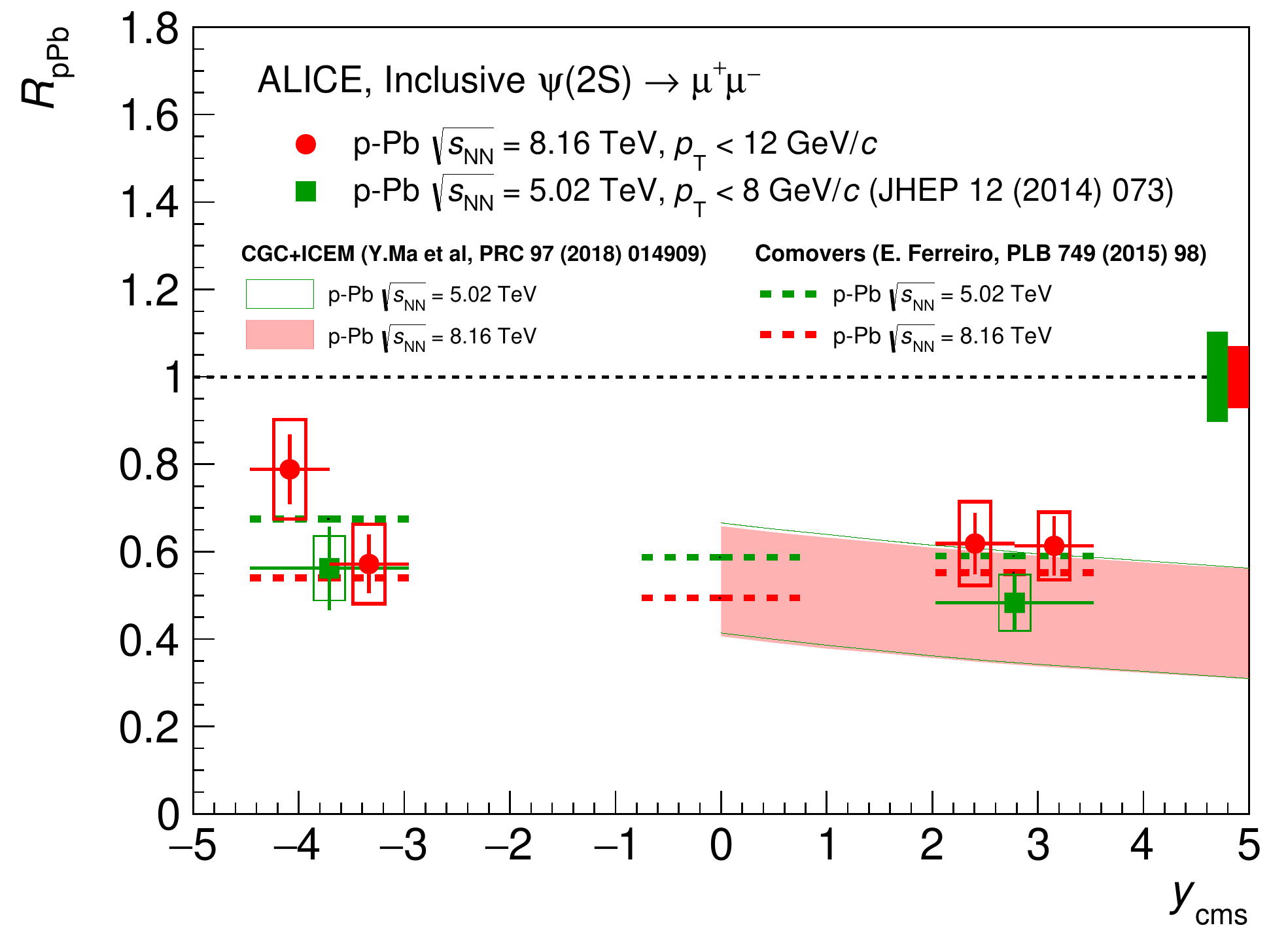}
\caption{Comparison of the rapidity dependence of $R_{\rm{pPb}}$ for \psip  in p--Pb collisions at $\sqrtSnn  = 8.16$ and 5.02 TeV~\cite{Abelev:2014zpa}. The error bars represent the statistical uncertainties, while the boxes correspond to uncorrelated systematic uncertainties and the boxes at $R_{\rm pPb}=1$ to correlated systematic uncertainties, separately shown for the two energies. The results are also compared with theoretical models that include final-state effects~\cite{Ma:2017rsu,Ferreiro:2014bia}. }
\label{fig:RAA_y_5_8_final}
\end{center}
\end{figure}

In Fig.~\ref{fig:RAA_y_5_8_final}, a comparison of the rapidity dependence of \psip suppression at $\sqrtSnn  = 8.16$ TeV and 5.02 TeV~\cite{Acharya:2018kxc} is presented, together with the corresponding results from theoretical models which implement final-state effects~\cite{Ferreiro:2014bia,Ma:2017rsu}. Both models fairly describe the \psip nuclear modification factor at both energies. The data at the two energies are in agreement within uncertainties. In Ref.~\cite{Abelev:2014zpa}, the reference for the \psip $R_{\rm pPb}$ evaluation at $\sqrtSnn = 5.02$ TeV was based only on the $\sqrt{s}=7$ TeV pp data available at that time~\cite{Abelev:2014qha}. If the procedure described in this paper would be adopted for the $\sqrtSnn = 5.02$ TeV result, the reference pp cross section would be lower by 12\% (corresponding to 0.9$\sigma$ on that quantity) and the $R_{\rm pPb}$ values would therefore be higher by the same amount.
In any case, the slightly stronger suppression predicted at $\sqrtSnn  = 8.16$ TeV and backward rapidity in Ref.~\cite{Ferreiro:2014bia,Albacete:2017qng}, related to the larger density of produced particles at higher energy, is beyond the sensitivity of the current measurement.

\begin{figure}[htbp]
\begin{center}     
\includegraphics[width=0.49\linewidth]{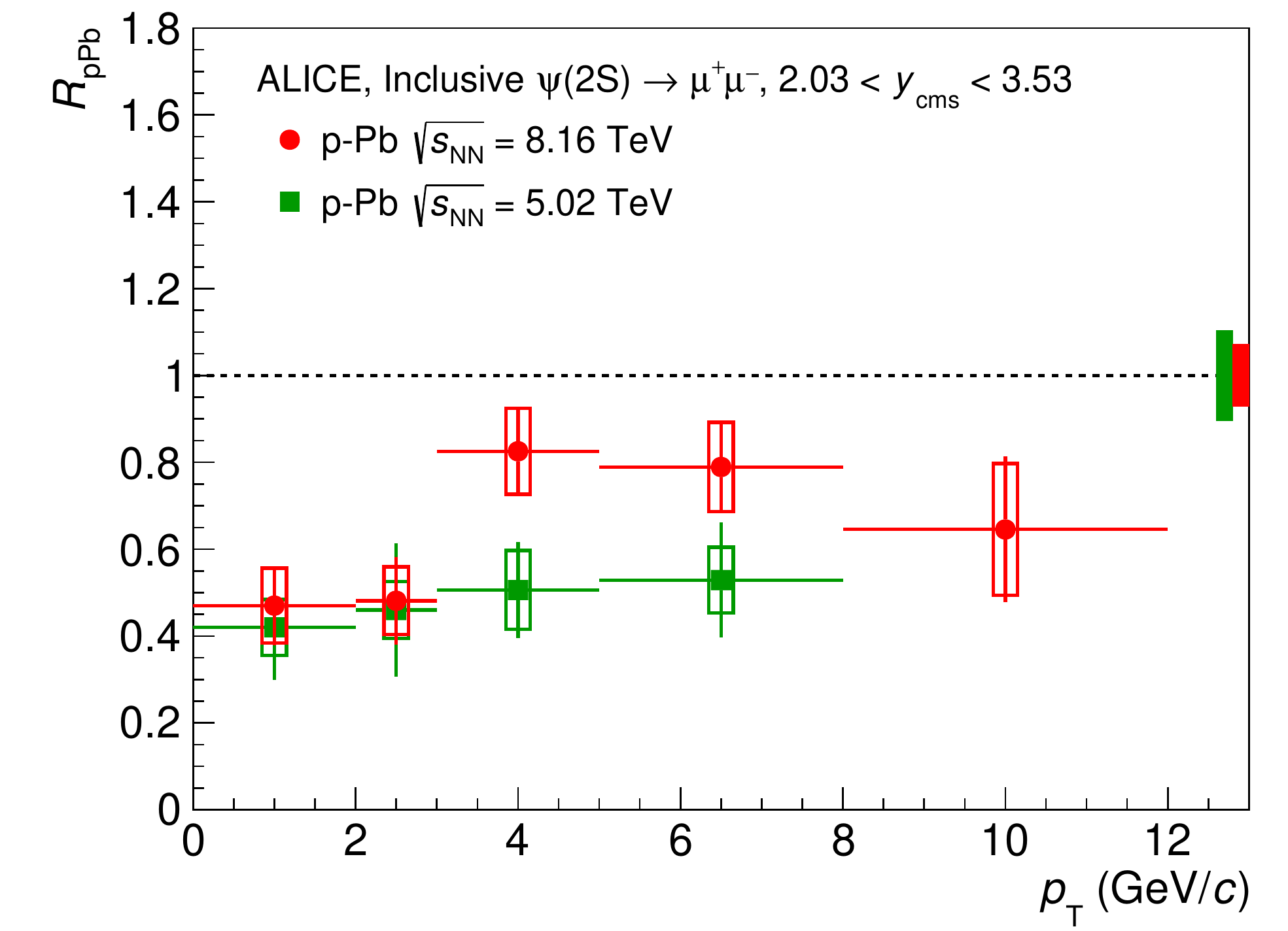}
\includegraphics[width=0.49\linewidth]{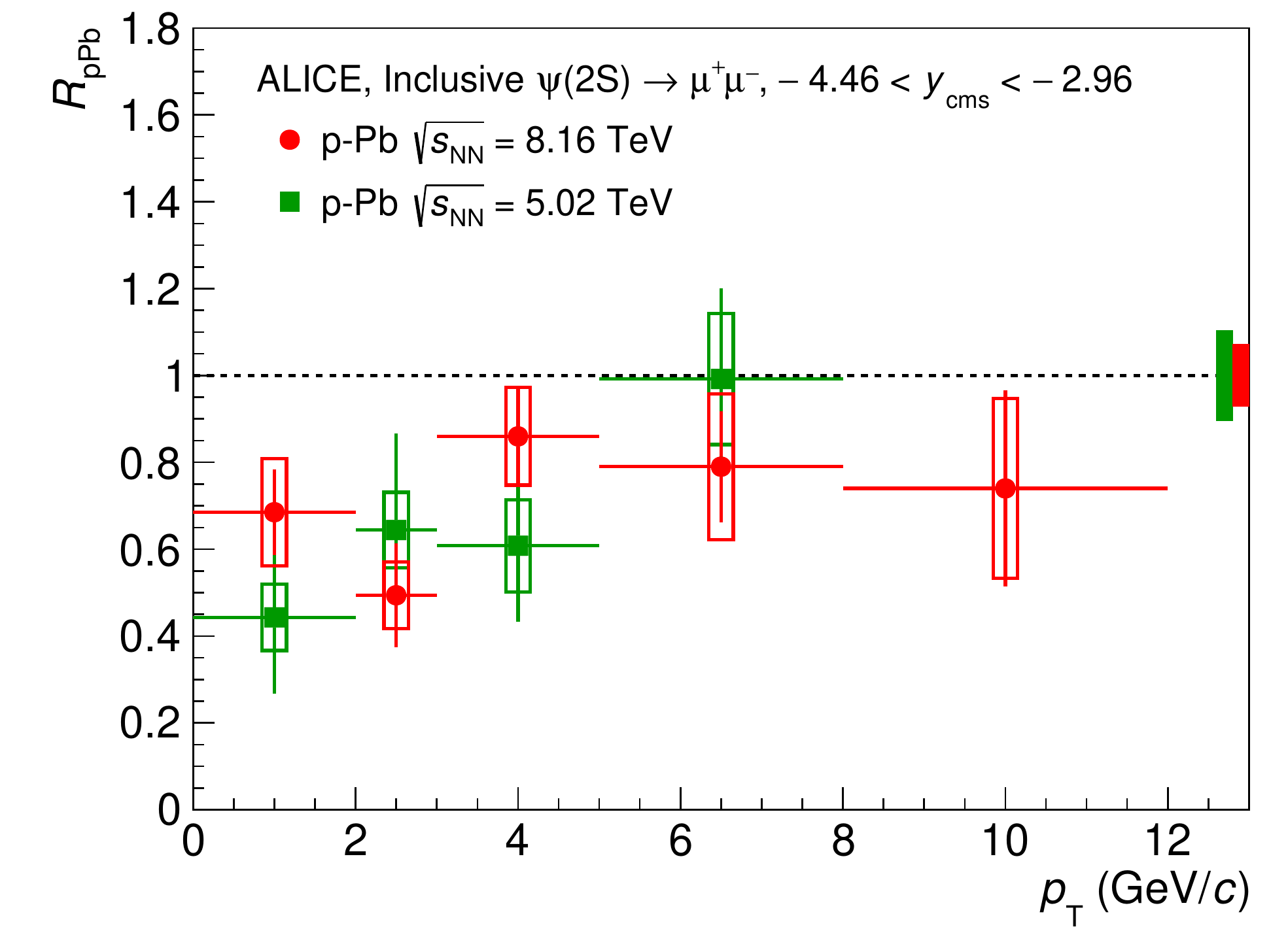}
\caption{Comparison of the transverse-momentum dependence of $R_{\rm{pPb}}$ for \psip in p--Pb collisions at $\sqrtSnn  = 8.16$ and 5.02 TeV~\cite{Abelev:2014zpa}. The error bars represent the statistical uncertainties, while the boxes correspond to uncorrelated systematic uncertainties  and the boxes at $R_{\rm pPb}=1$ to correlated systematic uncertainties, separately shown for the two energies.}
\label{fig:RAA_pt_5_8}
\end{center}
\end{figure}

In Fig.~\ref{fig:RAA_pt_5_8}, the results on the $p_{\rm T}$-dependence of $R_{\rm{pPb}}^{\psi(2{\rm S})}$ at the two energies studied by ALICE are presented. Within uncertainties, there is a fair agreement between the results, without a clear indication of a $p_{\rm T}$-dependence, except possibly for the  backward-rapidity results at $\sqrt{s_{\rm NN}}=5.02$ TeV which show a tendency to an increase at high $p_{\rm T}$. 

\begin{figure}[t!]
\begin{center}     
\includegraphics[width=0.6\linewidth]{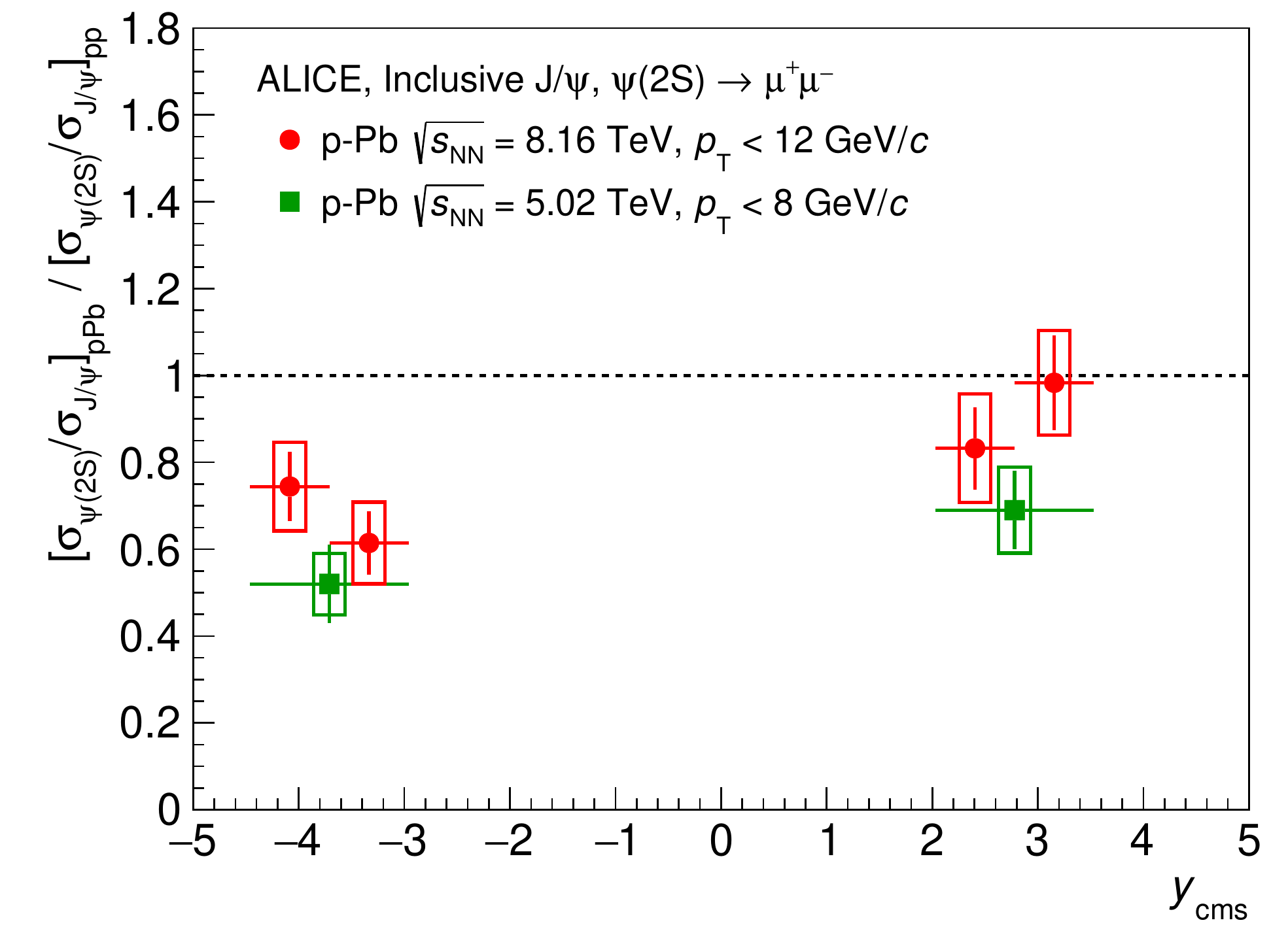}
\caption{Double ratio of $\psip$ and J/$\psi$ cross sections in \ppb and pp collisions as a function of rapidity, at $\sqrtSnn  = 8.16$ TeV, compared with the corresponding results at $\sqrtSnn  = 5.02$ TeV~\cite{Abelev:2014zpa}. The error bars represent the statistical uncertainties, while the boxes correspond to uncorrelated systematic uncertainties.}
\label{fig:dblratio}
\end{center}
\end{figure}

\begin{figure}[hbtp]
\begin{center}     
\includegraphics[width=0.49\linewidth]{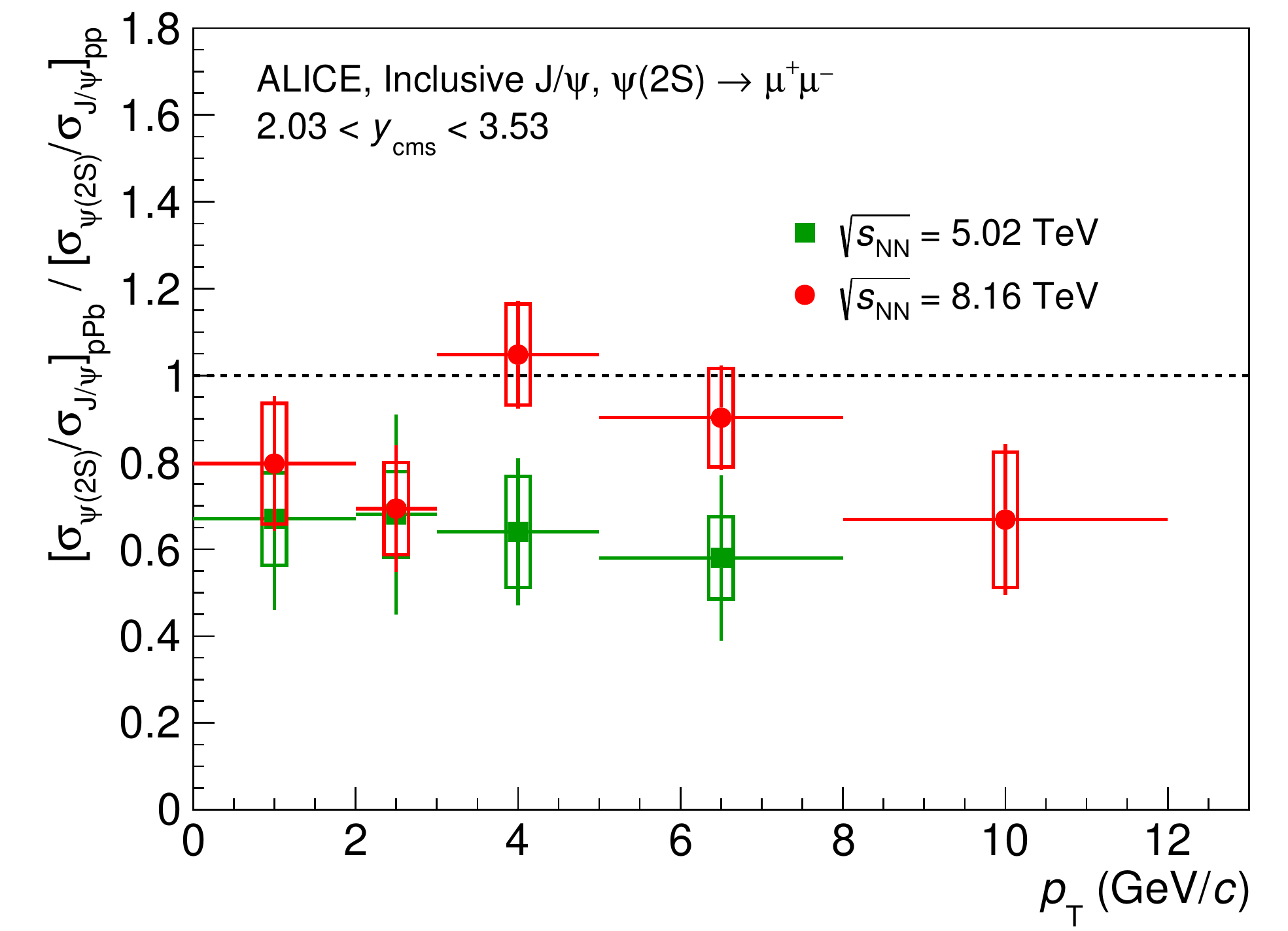}
\includegraphics[width=0.49\linewidth]{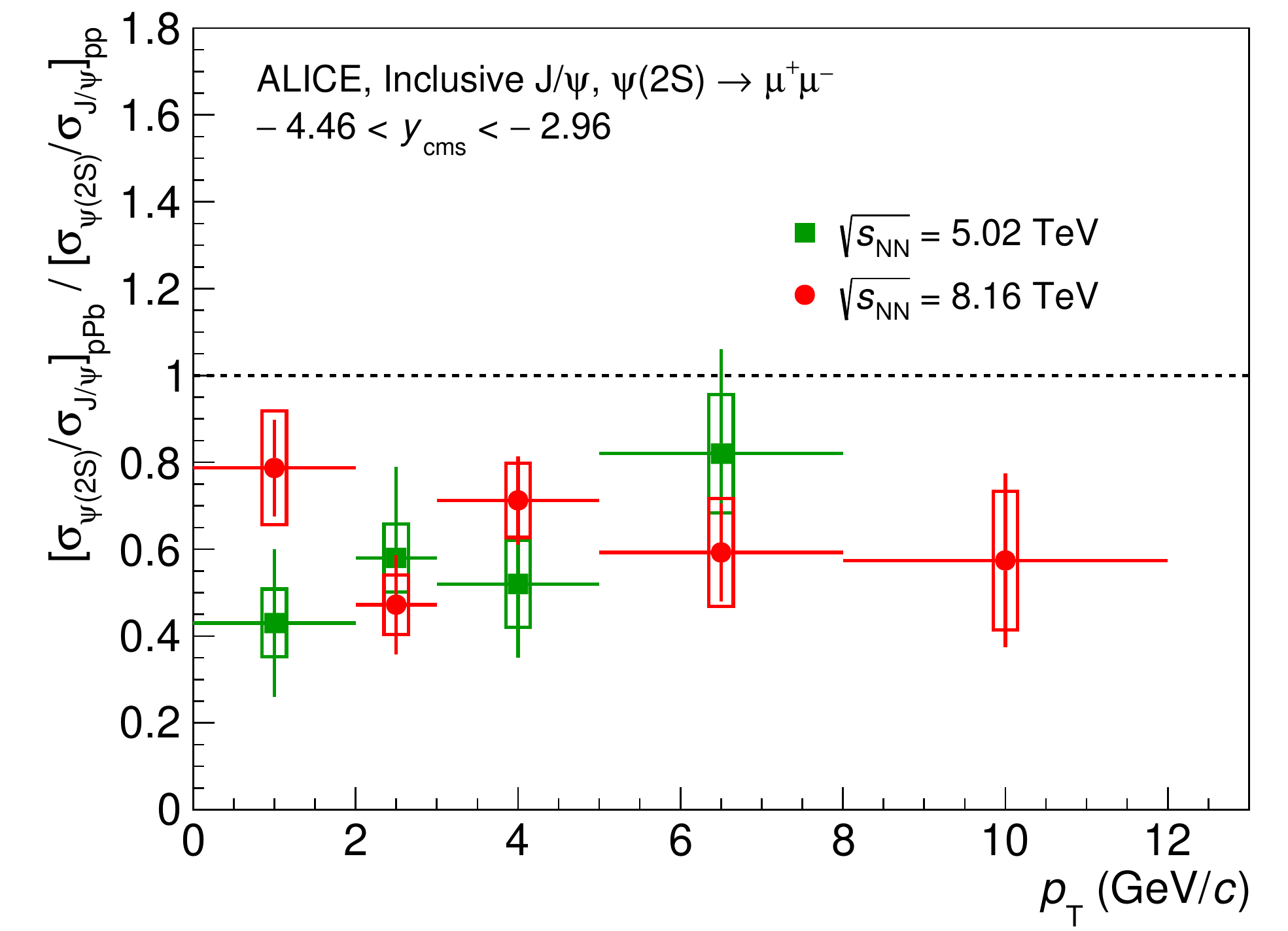}
\caption{Double ratio of $\psip$ and J/$\psi$ cross sections in \ppb and pp collisions as a function of transverse momentum, at forward (left) and backward (right) rapidity at $\sqrtSnn  = 8.16$ TeV, compared with the corresponding results at $\sqrtSnn  = 5.02$ TeV~\cite{Abelev:2014zpa}. The error bars represent the statistical uncertainties, while the boxes correspond to uncorrelated systematic uncertainties.}
\label{fig:dblratio_pt}
\end{center}
\end{figure}

Finally, also to ease comparisons with future results from other experiments, we present in Fig.~\ref{fig:dblratio}, as a function of $y_{\rm cms}$ and Fig.~\ref{fig:dblratio_pt}, as a function of $p_{\rm T}$, the values of the double ratio of the $\psip$ and J/$\psi$ cross sections between \ppb and pp. 
Clearly, these results confirm the features observed when comparing the nuclear modification factors for the two resonances, i.e., the $y_{\rm cms}$-dependence shows a relative suppression of the $\psip$ with respect to the J/$\psi$ at backward rapidity, while the $p_{\rm T}$-dependence does not indicate a clear trend.




\section{Conclusions}

The results of studies on the inclusive \psip production in p--Pb collisions at $\sqrtSnn = 8.16$ TeV, performed by ALICE, were shown. The data sample is about two times larger than the one at $\sqrtSnn = 5.02$ TeV, which was the object of a previous analysis~\cite{Acharya:2018kxc}.

The values of the nuclear modification factor indicate a 30--40\% \psip suppression at both forward and backward rapidity, with no significant transverse momentum dependence. When compared with the corresponding values for J/$\psi$, a similar suppression is found at forward rapidity, likely dominated by initial-state effects such as nuclear shadowing. At backward rapidity, the \psip suppression is significantly stronger than that of \jpsi. This effect is well reproduced by theoretical models that complement initial-state with final-state break-up effects, which should be more important for the loosely bound \psip state.
These results also confirm, with a better accuracy and extending the $p_{\rm T}$ reach, the previous observations carried out by ALICE in \ppb collisions at $\sqrtSnn = 5.02$ TeV.


\newenvironment{acknowledgement}{\relax}{\relax}
\begin{acknowledgement}
\section*{Acknowledgements}

The ALICE Collaboration would like to thank all its engineers and technicians for their invaluable contributions to the construction of the experiment and the CERN accelerator teams for the outstanding performance of the LHC complex.
The ALICE Collaboration gratefully acknowledges the resources and support provided by all Grid centres and the Worldwide LHC Computing Grid (WLCG) collaboration.
The ALICE Collaboration acknowledges the following funding agencies for their support in building and running the ALICE detector:
A. I. Alikhanyan National Science Laboratory (Yerevan Physics Institute) Foundation (ANSL), State Committee of Science and World Federation of Scientists (WFS), Armenia;
Austrian Academy of Sciences, Austrian Science Fund (FWF): [M 2467-N36] and Nationalstiftung f\"{u}r Forschung, Technologie und Entwicklung, Austria;
Ministry of Communications and High Technologies, National Nuclear Research Center, Azerbaijan;
Conselho Nacional de Desenvolvimento Cient\'{\i}fico e Tecnol\'{o}gico (CNPq), Financiadora de Estudos e Projetos (Finep), Funda\c{c}\~{a}o de Amparo \`{a} Pesquisa do Estado de S\~{a}o Paulo (FAPESP) and Universidade Federal do Rio Grande do Sul (UFRGS), Brazil;
Ministry of Education of China (MOEC) , Ministry of Science \& Technology of China (MSTC) and National Natural Science Foundation of China (NSFC), China;
Ministry of Science and Education and Croatian Science Foundation, Croatia;
Centro de Aplicaciones Tecnol\'{o}gicas y Desarrollo Nuclear (CEADEN), Cubaenerg\'{\i}a, Cuba;
Ministry of Education, Youth and Sports of the Czech Republic, Czech Republic;
The Danish Council for Independent Research | Natural Sciences, the VILLUM FONDEN and Danish National Research Foundation (DNRF), Denmark;
Helsinki Institute of Physics (HIP), Finland;
Commissariat \`{a} l'Energie Atomique (CEA), Institut National de Physique Nucl\'{e}aire et de Physique des Particules (IN2P3) and Centre National de la Recherche Scientifique (CNRS) and R\'{e}gion des  Pays de la Loire, France;
Bundesministerium f\"{u}r Bildung und Forschung (BMBF) and GSI Helmholtzzentrum f\"{u}r Schwerionenforschung GmbH, Germany;
General Secretariat for Research and Technology, Ministry of Education, Research and Religions, Greece;
National Research, Development and Innovation Office, Hungary;
Department of Atomic Energy Government of India (DAE), Department of Science and Technology, Government of India (DST), University Grants Commission, Government of India (UGC) and Council of Scientific and Industrial Research (CSIR), India;
Indonesian Institute of Science, Indonesia;
Centro Fermi - Museo Storico della Fisica e Centro Studi e Ricerche Enrico Fermi and Istituto Nazionale di Fisica Nucleare (INFN), Italy;
Institute for Innovative Science and Technology , Nagasaki Institute of Applied Science (IIST), Japanese Ministry of Education, Culture, Sports, Science and Technology (MEXT) and Japan Society for the Promotion of Science (JSPS) KAKENHI, Japan;
Consejo Nacional de Ciencia (CONACYT) y Tecnolog\'{i}a, through Fondo de Cooperaci\'{o}n Internacional en Ciencia y Tecnolog\'{i}a (FONCICYT) and Direcci\'{o}n General de Asuntos del Personal Academico (DGAPA), Mexico;
Nederlandse Organisatie voor Wetenschappelijk Onderzoek (NWO), Netherlands;
The Research Council of Norway, Norway;
Commission on Science and Technology for Sustainable Development in the South (COMSATS), Pakistan;
Pontificia Universidad Cat\'{o}lica del Per\'{u}, Peru;
Ministry of Science and Higher Education and National Science Centre, Poland;
Korea Institute of Science and Technology Information and National Research Foundation of Korea (NRF), Republic of Korea;
Ministry of Education and Scientific Research, Institute of Atomic Physics and Ministry of Research and Innovation and Institute of Atomic Physics, Romania;
Joint Institute for Nuclear Research (JINR), Ministry of Education and Science of the Russian Federation, National Research Centre Kurchatov Institute, Russian Science Foundation and Russian Foundation for Basic Research, Russia;
Ministry of Education, Science, Research and Sport of the Slovak Republic, Slovakia;
National Research Foundation of South Africa, South Africa;
Swedish Research Council (VR) and Knut \& Alice Wallenberg Foundation (KAW), Sweden;
European Organization for Nuclear Research, Switzerland;
Suranaree University of Technology (SUT), National Science and Technology Development Agency (NSDTA) and Office of the Higher Education Commission under NRU project of Thailand, Thailand;
Turkish Atomic Energy Agency (TAEK), Turkey;
National Academy of  Sciences of Ukraine, Ukraine;
Science and Technology Facilities Council (STFC), United Kingdom;
National Science Foundation of the United States of America (NSF) and United States Department of Energy, Office of Nuclear Physics (DOE NP), United States of America.    
\end{acknowledgement}

\bibliographystyle{utphys}  
\bibliography{draft.bib}


\newpage
\appendix
\section{The ALICE Collaboration}
\label{app:collab}

\begingroup
\small
\begin{flushleft}
S.~Acharya\Irefn{org141}\And 
D.~Adamov\'{a}\Irefn{org95}\And 
A.~Adler\Irefn{org74}\And 
J.~Adolfsson\Irefn{org81}\And 
M.M.~Aggarwal\Irefn{org100}\And 
G.~Aglieri Rinella\Irefn{org34}\And 
M.~Agnello\Irefn{org30}\And 
N.~Agrawal\Irefn{org10}\textsuperscript{,}\Irefn{org54}\And 
Z.~Ahammed\Irefn{org141}\And 
S.~Ahmad\Irefn{org16}
% ...
\And
\renewcommand\labelenumi{\textsuperscript{\theenumi}~}

\section*{Affiliation notes}
\renewcommand\theenumi{\roman{enumi}}
\begin{Authlist}
\item \Adef{org*}Deceased
\item \Adef{orgI}Italian National Agency for New Technologies, Energy and Sustainable Economic Development (ENEA), Bologna, Italy
\item \Adef{orgII}Dipartimento DET del Politecnico di Torino, Turin, Italy
\item \Adef{orgIII}M.V. Lomonosov Moscow State University, D.V. Skobeltsyn Institute of Nuclear, Physics, Moscow, Russia
\item \Adef{orgIV}Department of Applied Physics, Aligarh Muslim University, Aligarh, India
\item \Adef{orgV}Institute of Theoretical Physics, University of Wroclaw, Poland
\end{Authlist}

\section*{Collaboration Institutes}
\renewcommand\theenumi{\arabic{enumi}~}
\begin{Authlist}
\item \Idef{org1}A.I. Alikhanyan National Science Laboratory (Yerevan Physics Institute) Foundation, Yerevan, Armenia
\item \Idef{org2}Bogolyubov Institute for Theoretical Physics, National Academy of Sciences of Ukraine, Kiev, Ukraine
\item \Idef{org3}Bose Institute, Department of Physics  and Centre for Astroparticle Physics and Space Science (CAPSS), Kolkata, India
\item \Idef{org4}Budker Institute for Nuclear Physics, Novosibirsk, Russia
\item \Idef{org5}California Polytechnic State University, San Luis Obispo, California, United States
\item \Idef{org6}Central China Normal University, Wuhan, China
\item \Idef{org7}Centre de Calcul de l'IN2P3, Villeurbanne, Lyon, France
\item \Idef{org8}Centro de Aplicaciones Tecnol\'{o}gicas y Desarrollo Nuclear (CEADEN), Havana, Cuba
\item \Idef{org9}Centro de Investigaci\'{o}n y de Estudios Avanzados (CINVESTAV), Mexico City and M\'{e}rida, Mexico
\item \Idef{org10}Centro Fermi - Museo Storico della Fisica e Centro Studi e Ricerche ``Enrico Fermi', Rome, Italy
\item \Idef{org11}Chicago State University, Chicago, Illinois, United States
\item \Idef{org12}China Institute of Atomic Energy, Beijing, China
\item \Idef{org13}Comenius University Bratislava, Faculty of Mathematics, Physics and Informatics, Bratislava, Slovakia
\item \Idef{org14}COMSATS University Islamabad, Islamabad, Pakistan
\item \Idef{org15}Creighton University, Omaha, Nebraska, United States
\item \Idef{org16}Department of Physics, Aligarh Muslim University, Aligarh, India
\item \Idef{org17}Department of Physics, Pusan National University, Pusan, Republic of Korea
\item \Idef{org18}Department of Physics, Sejong University, Seoul, Republic of Korea
\item \Idef{org19}Department of Physics, University of California, Berkeley, California, United States
\item \Idef{org20}Department of Physics, University of Oslo, Oslo, Norway
\item \Idef{org21}Department of Physics and Technology, University of Bergen, Bergen, Norway
\item \Idef{org22}Dipartimento di Fisica dell'Universit\`{a} 'La Sapienza' and Sezione INFN, Rome, Italy
\item \Idef{org23}Dipartimento di Fisica dell'Universit\`{a} and Sezione INFN, Cagliari, Italy
\item \Idef{org24}Dipartimento di Fisica dell'Universit\`{a} and Sezione INFN, Trieste, Italy
\item \Idef{org25}Dipartimento di Fisica dell'Universit\`{a} and Sezione INFN, Turin, Italy
\item \Idef{org26}Dipartimento di Fisica e Astronomia dell'Universit\`{a} and Sezione INFN, Bologna, Italy
\item \Idef{org27}Dipartimento di Fisica e Astronomia dell'Universit\`{a} and Sezione INFN, Catania, Italy
\item \Idef{org28}Dipartimento di Fisica e Astronomia dell'Universit\`{a} and Sezione INFN, Padova, Italy
\item \Idef{org29}Dipartimento di Fisica `E.R.~Caianiello' dell'Universit\`{a} and Gruppo Collegato INFN, Salerno, Italy
\item \Idef{org30}Dipartimento DISAT del Politecnico and Sezione INFN, Turin, Italy
\item \Idef{org31}Dipartimento di Scienze e Innovazione Tecnologica dell'Universit\`{a} del Piemonte Orientale and INFN Sezione di Torino, Alessandria, Italy
\item \Idef{org32}Dipartimento di Scienze MIFT, Universit\`{a} di Messina, Messina, Italy
\item \Idef{org33}Dipartimento Interateneo di Fisica `M.~Merlin' and Sezione INFN, Bari, Italy
\item \Idef{org34}European Organization for Nuclear Research (CERN), Geneva, Switzerland
\item \Idef{org35}Faculty of Electrical Engineering, Mechanical Engineering and Naval Architecture, University of Split, Split, Croatia
\item \Idef{org36}Faculty of Engineering and Science, Western Norway University of Applied Sciences, Bergen, Norway
\item \Idef{org37}Faculty of Nuclear Sciences and Physical Engineering, Czech Technical University in Prague, Prague, Czech Republic
\item \Idef{org38}Faculty of Science, P.J.~\v{S}af\'{a}rik University, Ko\v{s}ice, Slovakia
\item \Idef{org39}Frankfurt Institute for Advanced Studies, Johann Wolfgang Goethe-Universit\"{a}t Frankfurt, Frankfurt, Germany
\item \Idef{org40}Fudan University, Shanghai, China
\item \Idef{org41}Gangneung-Wonju National University, Gangneung, Republic of Korea
\item \Idef{org42}Gauhati University, Department of Physics, Guwahati, India
\item \Idef{org43}Helmholtz-Institut f\"{u}r Strahlen- und Kernphysik, Rheinische Friedrich-Wilhelms-Universit\"{a}t Bonn, Bonn, Germany
\item \Idef{org44}Helsinki Institute of Physics (HIP), Helsinki, Finland
\item \Idef{org45}High Energy Physics Group,  Universidad Aut\'{o}noma de Puebla, Puebla, Mexico
\item \Idef{org46}Hiroshima University, Hiroshima, Japan
\item \Idef{org47}Hochschule Worms, Zentrum  f\"{u}r Technologietransfer und Telekommunikation (ZTT), Worms, Germany
\item \Idef{org48}Horia Hulubei National Institute of Physics and Nuclear Engineering, Bucharest, Romania
\item \Idef{org49}Indian Institute of Technology Bombay (IIT), Mumbai, India
\item \Idef{org50}Indian Institute of Technology Indore, Indore, India
\item \Idef{org51}Indonesian Institute of Sciences, Jakarta, Indonesia
\item \Idef{org52}INFN, Laboratori Nazionali di Frascati, Frascati, Italy
\item \Idef{org53}INFN, Sezione di Bari, Bari, Italy
\item \Idef{org54}INFN, Sezione di Bologna, Bologna, Italy
\item \Idef{org55}INFN, Sezione di Cagliari, Cagliari, Italy
\item \Idef{org56}INFN, Sezione di Catania, Catania, Italy
\item \Idef{org57}INFN, Sezione di Padova, Padova, Italy
\item \Idef{org58}INFN, Sezione di Roma, Rome, Italy
\item \Idef{org59}INFN, Sezione di Torino, Turin, Italy
\item \Idef{org60}INFN, Sezione di Trieste, Trieste, Italy
\item \Idef{org61}Inha University, Incheon, Republic of Korea
\item \Idef{org62}Institute for Nuclear Research, Academy of Sciences, Moscow, Russia
\item \Idef{org63}Institute for Subatomic Physics, Utrecht University/Nikhef, Utrecht, Netherlands
\item \Idef{org64}Institute of Experimental Physics, Slovak Academy of Sciences, Ko\v{s}ice, Slovakia
\item \Idef{org65}Institute of Physics, Homi Bhabha National Institute, Bhubaneswar, India
\item \Idef{org66}Institute of Physics of the Czech Academy of Sciences, Prague, Czech Republic
\item \Idef{org67}Institute of Space Science (ISS), Bucharest, Romania
\item \Idef{org68}Institut f\"{u}r Kernphysik, Johann Wolfgang Goethe-Universit\"{a}t Frankfurt, Frankfurt, Germany
\item \Idef{org69}Instituto de Ciencias Nucleares, Universidad Nacional Aut\'{o}noma de M\'{e}xico, Mexico City, Mexico
\item \Idef{org70}Instituto de F\'{i}sica, Universidade Federal do Rio Grande do Sul (UFRGS), Porto Alegre, Brazil
\item \Idef{org71}Instituto de F\'{\i}sica, Universidad Nacional Aut\'{o}noma de M\'{e}xico, Mexico City, Mexico
\item \Idef{org72}iThemba LABS, National Research Foundation, Somerset West, South Africa
\item \Idef{org73}Jeonbuk National University, Jeonju, Republic of Korea
\item \Idef{org74}Johann-Wolfgang-Goethe Universit\"{a}t Frankfurt Institut f\"{u}r Informatik, Fachbereich Informatik und Mathematik, Frankfurt, Germany
\item \Idef{org75}Joint Institute for Nuclear Research (JINR), Dubna, Russia
\item \Idef{org76}Korea Institute of Science and Technology Information, Daejeon, Republic of Korea
\item \Idef{org77}KTO Karatay University, Konya, Turkey
\item \Idef{org78}Laboratoire de Physique des 2 Infinis, Irène Joliot-Curie, Orsay, France
\item \Idef{org79}Laboratoire de Physique Subatomique et de Cosmologie, Universit\'{e} Grenoble-Alpes, CNRS-IN2P3, Grenoble, France
\item \Idef{org80}Lawrence Berkeley National Laboratory, Berkeley, California, United States
\item \Idef{org81}Lund University Department of Physics, Division of Particle Physics, Lund, Sweden
\item \Idef{org82}Nagasaki Institute of Applied Science, Nagasaki, Japan
\item \Idef{org83}Nara Women{'}s University (NWU), Nara, Japan
\item \Idef{org84}National and Kapodistrian University of Athens, School of Science, Department of Physics , Athens, Greece
\item \Idef{org85}National Centre for Nuclear Research, Warsaw, Poland
\item \Idef{org86}National Institute of Science Education and Research, Homi Bhabha National Institute, Jatni, India
\item \Idef{org87}National Nuclear Research Center, Baku, Azerbaijan
\item \Idef{org88}National Research Centre Kurchatov Institute, Moscow, Russia
\item \Idef{org89}Niels Bohr Institute, University of Copenhagen, Copenhagen, Denmark
\item \Idef{org90}Nikhef, National institute for subatomic physics, Amsterdam, Netherlands
\item \Idef{org91}NRC Kurchatov Institute IHEP, Protvino, Russia
\item \Idef{org92}NRC «Kurchatov Institute»  - ITEP, Moscow, Russia
\item \Idef{org93}NRNU Moscow Engineering Physics Institute, Moscow, Russia
\item \Idef{org94}Nuclear Physics Group, STFC Daresbury Laboratory, Daresbury, United Kingdom
\item \Idef{org95}Nuclear Physics Institute of the Czech Academy of Sciences, \v{R}e\v{z} u Prahy, Czech Republic
\item \Idef{org96}Oak Ridge National Laboratory, Oak Ridge, Tennessee, United States
\item \Idef{org97}Ohio State University, Columbus, Ohio, United States
\item \Idef{org98}Petersburg Nuclear Physics Institute, Gatchina, Russia
\item \Idef{org99}Physics department, Faculty of science, University of Zagreb, Zagreb, Croatia
\item \Idef{org100}Physics Department, Panjab University, Chandigarh, India
\item \Idef{org101}Physics Department, University of Jammu, Jammu, India
\item \Idef{org102}Physics Department, University of Rajasthan, Jaipur, India
\item \Idef{org103}Physikalisches Institut, Eberhard-Karls-Universit\"{a}t T\"{u}bingen, T\"{u}bingen, Germany
\item \Idef{org104}Physikalisches Institut, Ruprecht-Karls-Universit\"{a}t Heidelberg, Heidelberg, Germany
\item \Idef{org105}Physik Department, Technische Universit\"{a}t M\"{u}nchen, Munich, Germany
\item \Idef{org106}Politecnico di Bari, Bari, Italy
\item \Idef{org107}Research Division and ExtreMe Matter Institute EMMI, GSI Helmholtzzentrum f\"ur Schwerionenforschung GmbH, Darmstadt, Germany
\item \Idef{org108}Rudjer Bo\v{s}kovi\'{c} Institute, Zagreb, Croatia
\item \Idef{org109}Russian Federal Nuclear Center (VNIIEF), Sarov, Russia
\item \Idef{org110}Saha Institute of Nuclear Physics, Homi Bhabha National Institute, Kolkata, India
\item \Idef{org111}School of Physics and Astronomy, University of Birmingham, Birmingham, United Kingdom
\item \Idef{org112}Secci\'{o}n F\'{\i}sica, Departamento de Ciencias, Pontificia Universidad Cat\'{o}lica del Per\'{u}, Lima, Peru
\item \Idef{org113}St. Petersburg State University, St. Petersburg, Russia
\item \Idef{org114}Stefan Meyer Institut f\"{u}r Subatomare Physik (SMI), Vienna, Austria
\item \Idef{org115}SUBATECH, IMT Atlantique, Universit\'{e} de Nantes, CNRS-IN2P3, Nantes, France
\item \Idef{org116}Suranaree University of Technology, Nakhon Ratchasima, Thailand
\item \Idef{org117}Technical University of Ko\v{s}ice, Ko\v{s}ice, Slovakia
\item \Idef{org118}The Henryk Niewodniczanski Institute of Nuclear Physics, Polish Academy of Sciences, Cracow, Poland
\item \Idef{org119}The University of Texas at Austin, Austin, Texas, United States
\item \Idef{org120}Universidad Aut\'{o}noma de Sinaloa, Culiac\'{a}n, Mexico
\item \Idef{org121}Universidade de S\~{a}o Paulo (USP), S\~{a}o Paulo, Brazil
\item \Idef{org122}Universidade Estadual de Campinas (UNICAMP), Campinas, Brazil
\item \Idef{org123}Universidade Federal do ABC, Santo Andre, Brazil
\item \Idef{org124}University of Cape Town, Cape Town, South Africa
\item \Idef{org125}University of Houston, Houston, Texas, United States
\item \Idef{org126}University of Jyv\"{a}skyl\"{a}, Jyv\"{a}skyl\"{a}, Finland
\item \Idef{org127}University of Liverpool, Liverpool, United Kingdom
\item \Idef{org128}University of Science and Technology of China, Hefei, China
\item \Idef{org129}University of South-Eastern Norway, Tonsberg, Norway
\item \Idef{org130}University of Tennessee, Knoxville, Tennessee, United States
\item \Idef{org131}University of the Witwatersrand, Johannesburg, South Africa
\item \Idef{org132}University of Tokyo, Tokyo, Japan
\item \Idef{org133}University of Tsukuba, Tsukuba, Japan
\item \Idef{org134}Universit\'{e} Clermont Auvergne, CNRS/IN2P3, LPC, Clermont-Ferrand, France
\item \Idef{org135}Universit\'{e} de Lyon, Universit\'{e} Lyon 1, CNRS/IN2P3, IPN-Lyon, Villeurbanne, Lyon, France
\item \Idef{org136}Universit\'{e} de Strasbourg, CNRS, IPHC UMR 7178, F-67000 Strasbourg, France, Strasbourg, France
\item \Idef{org137}Universit\'{e} Paris-Saclay Centre d'Etudes de Saclay (CEA), IRFU, D\'{e}partment de Physique Nucl\'{e}aire (DPhN), Saclay, France
\item \Idef{org138}Universit\`{a} degli Studi di Foggia, Foggia, Italy
\item \Idef{org139}Universit\`{a} degli Studi di Pavia, Pavia, Italy
\item \Idef{org140}Universit\`{a} di Brescia, Brescia, Italy
\item \Idef{org141}Variable Energy Cyclotron Centre, Homi Bhabha National Institute, Kolkata, India
\item \Idef{org142}Warsaw University of Technology, Warsaw, Poland
\item \Idef{org143}Wayne State University, Detroit, Michigan, United States
\item \Idef{org144}Westf\"{a}lische Wilhelms-Universit\"{a}t M\"{u}nster, Institut f\"{u}r Kernphysik, M\"{u}nster, Germany
\item \Idef{org145}Wigner Research Centre for Physics, Budapest, Hungary
\item \Idef{org146}Yale University, New Haven, Connecticut, United States
\item \Idef{org147}Yonsei University, Seoul, Republic of Korea
\end{Authlist}
\endgroup
\end{document}